\documentclass[12pt,preprint]{aastex} 
\usepackage[left=1in,top=1in,bottom=1in,right=1in]{geometry}
\usepackage{natbib}
\usepackage{lscape}
\newcommand       \be		{\begin{equation}}
\newcommand       \ee		{\end{equation}}

\newcommand       \kpc		{\,{\rm kpc \,}}
\newcommand       \pc		{\,{\rm pc \,}}

\newcommand       \kb		{\,{\rm k_b \,}}
\newcommand       \yr		{\,{\rm yr \,}}
\newcommand       \Myr		{\,{\rm Myr \,}}
\newcommand       \cm		{\,{\rm cm \,}}
\newcommand       \s		{\,{\rm s \,}}

\newcommand \kms {\,{\rm km \,\, s}^{-1}}

\begin{document}

\title{Milky Way Star Forming Complexes and the Turbulent Motion of the Galaxy's Molecular Gas}
\author{Eve J. Lee\altaffilmark{1,2}, Norman Murray\altaffilmark{2,3}, Mubdi Rahman\altaffilmark{1}}
\altaffiltext{1}{Department of Astronomy and Astrophysics, University of
Toronto, 50 St. George Street, Toronto, ON M5S 3H4, Canada; elee@astro.utoronto.ca, rahman@astro.utoronto.ca}
\altaffiltext{2}{Canadian Institute for Theoretical Astrophysics, 60 St.
George Street, University of Toronto, Toronto ON M5S 3H8,
Canada; elee@cita.utoronto.ca, murray@cita.utoronto.ca}
\altaffiltext{3}{Canada Research Chair in Astrophysics}

\begin{abstract}
We analyze Spitzer GLIMPSE, MSX, and WMAP images of the Milky Way to identify 8 micron and free-free sources in the Galaxy. Seventy-two of the eighty-eight WMAP sources have coverage in the GLIMPSE and MSX surveys suitable for identifying massive star forming complexes (SFC). We measure the ionizing luminosity functions of the SFCs and study their role in the turbulent motion of the Galaxy's molecular gas. We find a total Galactic free-free flux $f_{\nu}$ = 46177.6 Jy; the 72 WMAP sources with full 8 micron coverage account for 34263.5 Jy ($\sim 75\%$), with both measurements made at $\nu$=94GHz (W band). We find a total of 280 SFCs, of which 168 have unique kinematic distances and free-free luminosities. We use a simple model for the radial distribution of star formation to estimate the free-free and ionizing luminosity for the sources lacking distance determinations. The total dust-corrected ionizing luminosity is $Q = 2.9 \pm 0.5~\rm{x}~10^{53}~{\rm photons}~s^{-1}$, which implies a galactic star formation rate of $\dot{M}_{*} = 1.2 \pm 0.2~M_{\sun}~\rm{yr}^{-1}$. We present the (ionizing) luminosity function of the SFCs, and show that 24 sources emit half the ionizing luminosity of the Galaxy. The SFCs appear as bubbles in GLIMPSE or MSX images; the radial velocities associated with the bubble walls allow us to infer the expansion velocity of the bubbles. We calculate the kinetic luminosity of the bubble expansion and compare it to the turbulent luminosity of the inner molecular disk. SFCs emitting $80\%$ of the total galactic free-free luminosity produce a kinetic luminosity equal to $65\%$ of the turbulent luminosity in the inner molecular disk. This suggests that the expansion of the bubbles is a major driver of the turbulent motion of the inner Milky Way molecular gas. 
\end{abstract}

\keywords{infrared: ISM -- ISM: \ion{H}{2} regions -- stars: formation}

\section{Introduction}
\label{sec:intro}
Turbulence is ubiquitous in astrophysical fluids, from the intergalactic medium of galaxy clusters through the interstellar medium of galaxies to protostellar accretion disks.  Observations of CO linewidths corresponding to supersonic velocity dispersions have led to the suggestion that turbulence in molecular clouds may slow gravitational contraction and limit the rate of star formation, e.g., \citet{shu}. If the motions are turbulence-dominated, there must be a driving energy source, since undriven turbulence decays on the order of a crossing time \citep{1974ApJ...189..441G,maclow}. Currently proposed sources can be divided into two groups: gravitational energy released by accretion onto or through the galactic disk \citep[e.g.,][]{klessen10}, and nuclear energy. Gravitational energy can generate turbulence via instabilities such as bars or spiral arms~\citep{li05} on large scales, and protostellar outflows~\citep{li06, wang10} on small scales, while nuclear energy couples to the interstellar medium (ISM) turbulence via expanding \ion{H}{2} regions~\citep{matzner02}, radiation pressure \citep{mqt05}, or supernovae~\citep{maclow04}. In the absence of star formation activities such as in the extended \ion{H}{1} disk, magnetorotational instability is proposed as one of the sources of turbulent motion \citep[e.g.][]{sellwood99}. None of the proposed mechanisms, however, have strong observational support.

Recently, \citet[][hereafter MR10]{paper1} identified 88 {\em Wilkinson Microwave Anisotropy Probe} (WMAP) galactic free-free sources in an effort to determine the galactic star formation rate (SFR). These WMAP sources contain bubble-like structures with radii ranging from 5 to 100 pc. Since the 18 most luminous sources identified produced over half the total ionizing luminosity of the Galaxy, they proposed that these sources contain over half the Galactic O star population, most likely in massive clusters\footnote{A candidate OB association has been found in G298 and a follow-up spectroscopic analysis confirms 15 O-type and 3 luminous blue variables or Wolf-Rayet stars~\citep{candidate, rahman11}.}. The ionizing photons produced by the clusters leak out of their host bubbles and ionize the surrounding area. These leaking photons create the extended low density (ELD) regions, originally described by \citet{mezger} as the dominant source of galactic radio free-free emission; the WMAP sources are argued to be the ELDs.

Noting the existence of numerous known \ion{H}{2} regions on the shells or walls of the bubbles in the WMAP sources, \citet[][hereafter RM10]{paper2} identified 40 star forming complexes (SFCs) in the 13 most luminous WMAP sources, responsible for one-third of the star formation in the Milky Way. The kinetic luminosity due to the expansion of the bubbles associated with the SFCs can account for a third of the turbulent luminosity in the Milky Way's molecular gas disk, suggesting that the bubble expansion is the primary driver of turbulence. They also found that the initial driver of the bubbles was not supernovae, as the regions were selected by their free-free luminosity; the free-free luminosity of a star cluster fades rapidly after the first supernovae explode, because the most massive stars dominate the free-free emission. Since the SFCs selected by RM10 are in the most luminous WMAP sources, further investigation is required to determine if the relationship between free-free and kinetic luminosity scales to the less luminous sources.

In this paper, we perform a follow-up study by investigating the remaining 75 WMAP sources and revisiting the previously-studied 13. In $\S$\ref{sec:data}, we describe the selection of the SFCs, discuss their expansion velocities, distance and free-free flux determination, and show comparison statistics to RM10. In $\S$\ref{sec:results}, we refine the galactic ionizing luminosity and star formation rate, present an empirical relationship between the 8$\mu$m and free-free emission, and show that the expansion of the SFCs is a significant driving mechanism of turbulent motion of the molecular gas inside the solar circle. We discuss the results in \S \ref{sec:discuss}. In $\S$\ref{sec:concl}, we summarize the results.

\section{Data Analysis}
\label{sec:data}
\subsection{Selection of Star Forming Complexes}
\label{ssec:selsfc}
MR10 identified 88 galactic free-free sources in the WMAP maximum entropy method free-free foreground emission map: 87 in the Galaxy, the 88th being the LMC. The 13 most luminous WMAP sources were then analyzed by RM10 using radial velocity measurements taken from the literature, combined with the morphological features in 8$\mu$m images. Here, we investigate all the WMAP free-free sources to obtain a complete sample of the Galactic free-free emitting SFCs.

In order to encompass an entire WMAP source in a single image, band 4 (8 $\mu$m) mosaics from the {\em Spitzer} Galactic Legacy Infrared Midplane Survey Extraordinaire (GLIMPSE) are stitched together using the Montage package. Forty-nine sources do not have sufficient GLIMPSE coverage; for these sources, {\em Midcourse Space Experiment} (MSX) band A images, also at 8 $\mu$m but with lower sensitivity and spatial resolution, are used instead. Visual inspection of the 8$\mu$m images revealed bubble-like morphologies outlined by bright regions along the walls that are well correlated with \ion{H}{2} regions found from SIMBAD.  Catalogues of radio recombination line (RRL) measurements taken from the literature are used to find the radial velocities of these \ion{H}{2} regions. In cases where there are no RRL measurement, H$\alpha$ or molecular line velocities are used (non-RRL velocities are used only for SFC selection and are excluded in further analysis). SFCs are selected based on the visual inspection of the shell morphology in the 8$\mu$m images, and a small range in RRL velocities, where we set the maximum spread to be $15\kms$, based on the theoretical study of expanding bubbles by~\citet{harper-clark}. Thus, we interpret the dispersion in the velocities of the \ion{H}{2} regions as the expansive motions of the bubbles inside the SFC.

\begin{figure}
\plotone{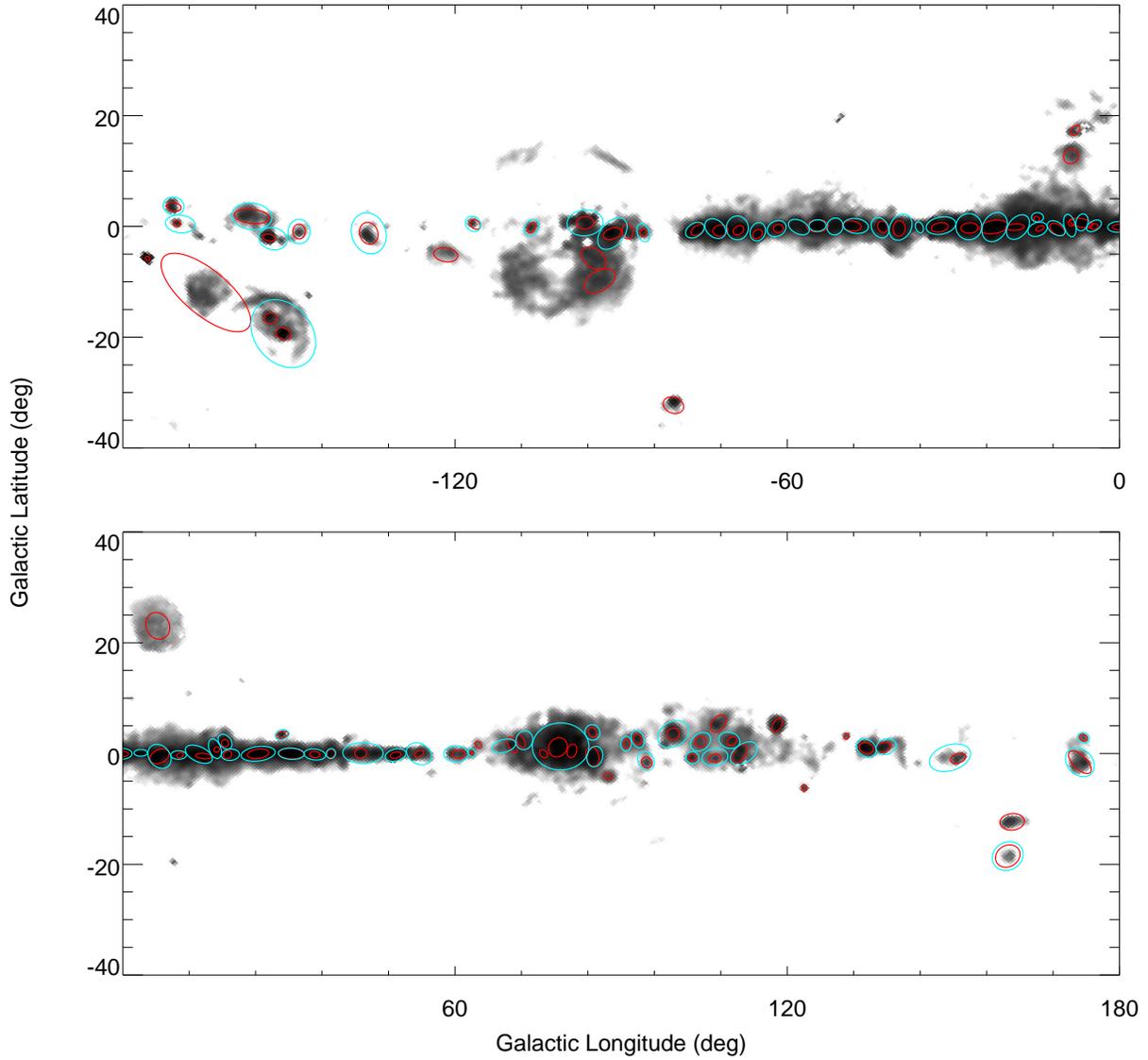}
\caption{\label{wmapsrc} Histogram-equalized image of all pixels above 0.02Jy in W band at 94 GHz from the WMAP free-free foreground emission map with the WMAP sources overlaid~(MR10). Original positions and sizes are indicated by red ellipses and the resized versions are indicated by turquoise ellipses.}
\end{figure}

Often 2 or 3 additional SFCs are found just outside the WMAP ellipses from MR10; in these cases, the WMAP source is either resized and reoriented to encompass all the observed SFCs, or new WMAP sources are created if RRL velocities of SFCs outside the original WMAP ellipses are significantly different from those inside. On the other hand, some SFCs are found to lie across multiple WMAP sources; in these cases, WMAP sources are merged appropriately. The mergers effectively eliminate three MR10 sources while the splits add 4 new sources. Both the original and the new WMAP sources are shown in Figure \ref{wmapsrc}. 

MR10 found 87 free-free sources in the Galaxy. After merging and splitting sources, we have 88 in which we can potentially find star forming complexes. However, ten of our WMAP sources do not have any 8$\mu$m coverage and are excluded from our SFC search. After the further exclusion of the WMAP regions with only partial $8\mu$m coverage (G64, G107, G118) or a lack of bubble detection (G87, G130, G271), a total of 72 WMAP sources are suitable for finding SFCs. The total free-free flux of the 72 WMAP sources is 34263.5 Jy, which is 74\% of the total Galactic free-free flux 46177.6 Jy. 

\begin{figure}
\plotone{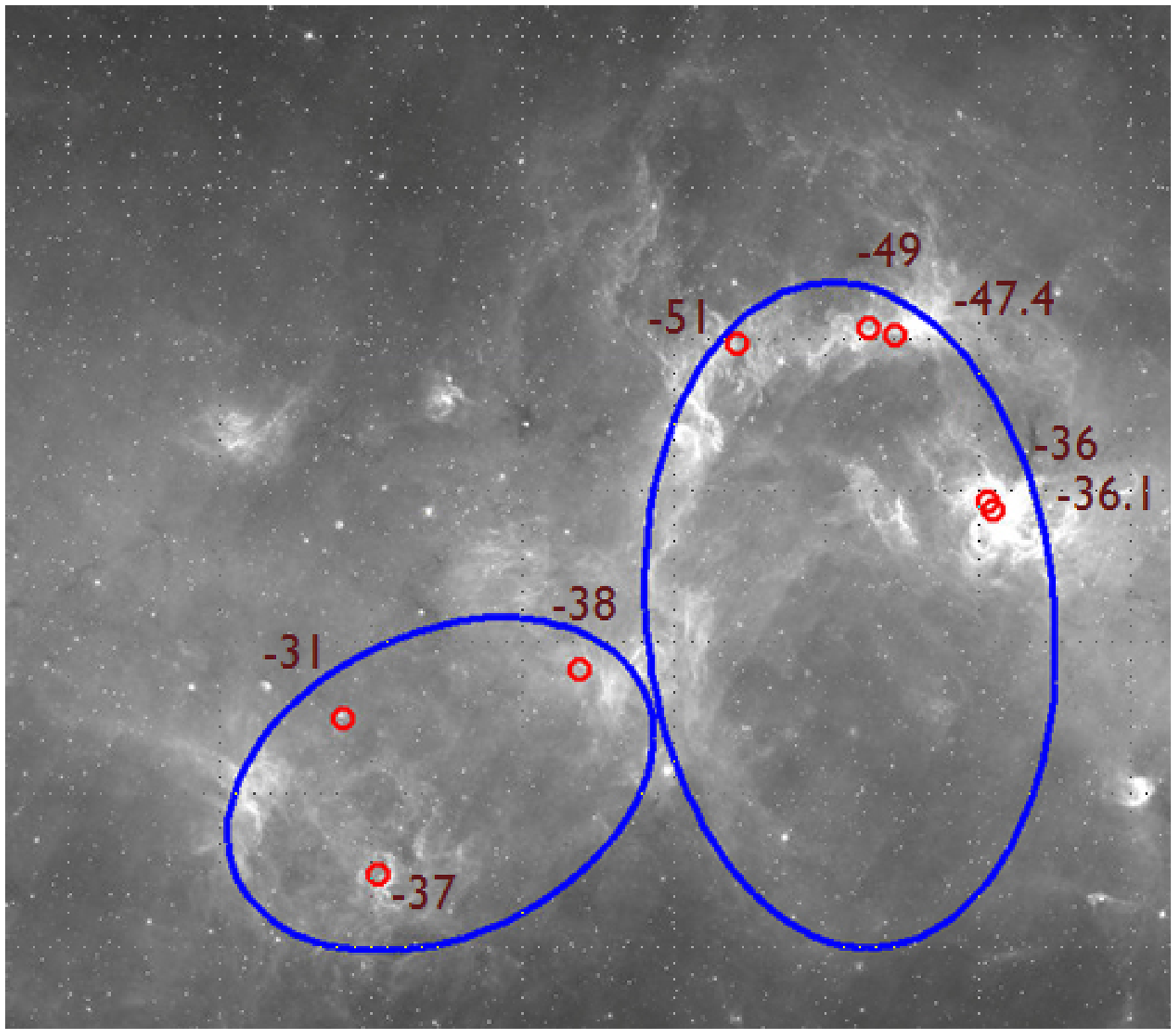}
\caption{\label{g317} Spitzer GLIMPSE image of a part of the G317 WMAP source with SFCs indicated by blue ellipses and the \ion{H}{2} region velocities indicated by red circles.}
\end{figure}

We find a total of 280 SFCs in these 72 sources; a sample selection is shown in Figure \ref{g317}. 

We note that MR10 estimate the total galactic free-free flux as 54211.6 Jy, measured in Ka band (33GHz). By summing up all the pixels in each of the 5 WMAP band images and plotting as a function of the frequency, we find $F_{\nu} \propto \nu^{-0.16 \pm 0.11}$, using a $12\%$ error as estimated from the H$\alpha$ emission discrepancy~\citep{wmap_fg}. This is consistent with the frequency dependence of free-free emission, due primarily to the variation in the Gaunt factor (see equation \ref{eq:ffem}). It is this weak frequency dependence which leads to the lower free-free flux quoted in this paper. All the free-free fluxes quoted in this paper are measured in the W band at 94 GHz from the WMAP free-free foreground emission map~(MR10).

\subsection{Properties of Star Forming Complexes}
\label{ssec:prop_sfc}

\subsubsection{Radial Velocities and Distances}
\label{sssec:vel_dist}
We interpret the morphology seen in the 8$\mu$m images as the result of quasi-spherical expansion, assuming that the RRL velocities sample the expansion rate of the bubble wall. However, due to the line-of-sight projection of the three-dimensional expansion motion and limited sampling of RRL velocities within the SFCs, a geometric correction is required to convert the measured velocity spread into an expansion velocity. The method outlined in RM10 is used, in which the measured half-spread velocity (defined as $(v_{max}-v_{min})/2$) is multiplied by a geometric factor found by running a Monte Carlo simulation where each velocity measurement is de-projected by assigning a random pair of spherical coordinate angles. The mean correction factor is 2.0, 1.4 and 1.2 for 5, 10 and 15 velocity measurements respectively~(RM10).

Defining the median of the local standard of rest (LSR) velocities of the \ion{H}{2} (RRL) regions as the velocity of the host SFC, a kinematic distance is determined by fitting the galactic longitude of the centroid of the SFC and the LSR velocity of the SFC using the~\citet{clemens85} rotation curve, assuming that the distance from the sun to the Galactic center is $R_{\sun}$ = 8.5$\kpc$, and a circular velocity $\theta = 220\kms$. The kinematic distance of SFCs with highly peculiar (forbidden) velocities are re-calculated allowing for a deviation in the velocity of $\pm$ 20$\kms$. 

SFCs that lie in the inner galaxy ($r<R_{\sun}$) suffer from the kinematic distance ambiguity (KDA); the source could lie at either the near or far points where the projected line of sight velocity matches the measured velocity. Absorption line data from the literature are used to break the degeneracy. We use absorption line data from a number of sources. First, \citet{kda_BUcat} provide both \ion{H}{1} emission/absorption and self-absorption data, and a corresponding estimate of the reliability of the result. In case these two measurements do not agree, we use the result with the higher assigned confidence level. In case the methods disagree and are given the same level of confidence, we compare their quoted distance to our distance derived from RRL velocities. 

Other sources from the radial velocity literature that we employ include \citet{kda_busfield}, who use \ion{H}{1} self-absorption data while \citet{kda_sewilo} use $\rm{H}_{2}\rm{CO}$ absorption data. Both \citet{kda_fish} and \citet{kda_kolpak} provide \ion{H}{1} line absorption (measured against \ion{H}{2} continuum background). \citet{vrad_caswell_haynes} use a combination of \ion{H}{1} and $\rm{H}_{2}\rm{CO}$ absorption data as well as optical counterparts to the \ion{H}{2} regions. 

\citet{kda_palagi} and \citet{vrad_gill} quote only the distances (and not the radial velocity $v_r$), where the former resolve the KDA by comparing with the result of \citet{kda_downes} and \citet{solomon87} and the latter by using $\rm{H}_{2}\rm{CO}$ absorption data and the spectrophotometry of associated stars. In these and similar cases, when only the absorption line-based distance is given (and not $v_r$), we compare the distance in the literature to the distance determined from the radio recombination line $v_r$ combined with the Clemens rotation curve, and choose the closest match. 

Given a SFC with multiple resolutions of the KDA, we assume majority rules; for example, if a SFC has 5 KDA measurements with 3 near and 2 far, we take the near distance. In case of a tie, we take the measurement with (literature reported) higher confidence. Failing that, we check the vertical distance of the SFC from the plane and take the near distance if the far distance gives the height greater than 100pc. If the ambiguity remains, we discard measurements from regions that do not project onto the shell walls in $8\mu$m images. Finally, we check the physical size of the SFC---based on the hypothesis that the bubbles are blown by massive stars, SFC size can be no bigger than the expansion velocity times the lifetime of massive O stars (4 Myr). There are 26 SFCs with $\tau_{dyn}$ = physical radius/ expansion velocity $>$ 4 Myr. Of these 26 SFCs, 7 of them do not have corrected mean-spread velocity. As the median discrepancy of the dynamical time is $\sim 49\%$, not a factor of 2 or higher, and the physical radius of the bubble is on the order of or less than 100 pc, we interpret them as an underestimate of expansion velocity.

Three SFCs are assigned a kinematic distance manually: G43 SFC92, G359 SFC0, G359 SFC1. G43 SFC92 is assigned $6\kpc$ from the tangent point measurement of \citet{kda_BUcat} (the measured near and far distance are $5\kpc$ and $7\kpc$ respectively). G359 SFC0 and SFC1 are both assumed to lie at $8\kpc$ as they are most likely associated with Arches cluster and Quintuplet cluster, respectively.

Use of the Galactic rotation curve as a distance estimator fails near the Galactic centre, giving erroneously far or near distance for SFCs with small Galactic longitude. As will be described in $\S$\ref{sssec:spiral_arm}, we can infer the association of SFCs with the $\sim3\kpc$ molecular ring \citep[see, eg.,][]{dame01} using only Galactic longitude and LSR velocity. If a SFC we expect to be part of the molecular ring is assigned a distance larger than $14\kpc$, we manually change the distance to $12\kpc$; conversely, if the SFC is assigned to a distance closer than $3\kpc$, we manually change the distance to $4\kpc$. 

A total of 194 SFCs require KDA resolution; we find unique distances for 113 of these 194 sources. Combined with the SFCs that do not require the resolution, we identify a unique distance estimate to 168 out of our 280 SFCs.


\citet{reid09}, based on their parallax study, suggest that $\theta=250\kms$ rather than $220\kms$. Assuming that \citet{reid09} is correct, using the distance calculated by the \citet{clemens85} curve results in outer galaxy distances being, on average, overestimated by a factor of $1.5$. The inner galaxy source distances are much less sensitive to $\theta$, changing by approximately 1\% on average. \citet{mcmillan10} point out that $R_{\sun}$ and $\theta$ cannot be reliably determined and are highly sensitive to model parameters; furthermore, they point out that the result of \citet{reid09} may not be accurate due to the underestimated peculiar motion of the Sun with respect to the LSR. Nevertheless, our results shows that the distance to the inner galaxy SFCs is largely insensitive to the value of $\theta$, and the distance discrepancy in the outer sources, as will be described in $\S$~\ref{sssec:lum_SFR}, can alter our SFR by $\sim6\%$.

\subsubsection{Free-free fluxes}
\label{sssec:ff-fluxes}
The WMAP beam is significantly larger than the typical SFC radius, resulting in spatial confusion between the free-free emission of different SFCs. To overcome this confusion and attribute free-free flux estimates to individual SFCs, we divide the total WMAP flux among the corresponding SFCs by the relative $8\mu$m flux. (Here we define the total WMAP free-free flux as the isophotal flux calculated in MR10). The relationship between free-free flux and $8\mu$m emission is discussed in $\S$\ref{ssec:ff8mrel} where it is shown that $8\mu$m is a good tracer of free-free emission. Previously, \citet{gmc} used relative RRL flux to divide up total WMAP free-free flux. These two methods agree well, with correlation coefficient $\sim$0.7. We prefer using the $8\mu$m flux because it does not require  uniform sampling (both in number and wavelength) of RRL measurements among the SFCs in the same host WMAP, sampling which is rarely available. All 280 SFCs are assigned a free-free flux using this method. As a result, we find the total flux of the SFCs to be 34234.8 Jy (the $0.1\%$ difference from the total 72 WMAP source free-free flux appears to be due to rounding error in our python scripts), leaving $26\%$ of the total Galactic free-free flux 46177.6 Jy unaccounted for. Ultimately, 166 SFCs have both unique kinematic distance and free-free flux determinations.

\subsubsection{Radial velocity and distance uncertainties due to streaming motions of \ion{H}{2} gas}
\label{sssec:streaming}
It is possible that the \ion{H}{2} region line-of-sight velocities may probe the superposition of the streaming motion of the ionized gas driven locally by the O stars powering the free-free emission \citep[$\sim 10~\kms$][]{matzner02}, in combination with the expansion of the bubble. However, Table \ref{sfclist} shows that the average dynamical time of the SFCs $\tau_{dyn}\equiv R/\Delta v_c$ is 4 Myr. This is roughly the lifetime of the most massive O stars, which dominate the production of ionizing radiation. In other words, the bubble expansion velocity, calculated from the bubble size divided by the upper limit for the age of the system is of order $10~\kms$. This is consistent with the independent measurement of the expansion velocity provided by the RRL data: the mean expansion velocity of all the SFCs is found to be $12~\pm~8~\kms$. 

\begin{figure}
\plotone{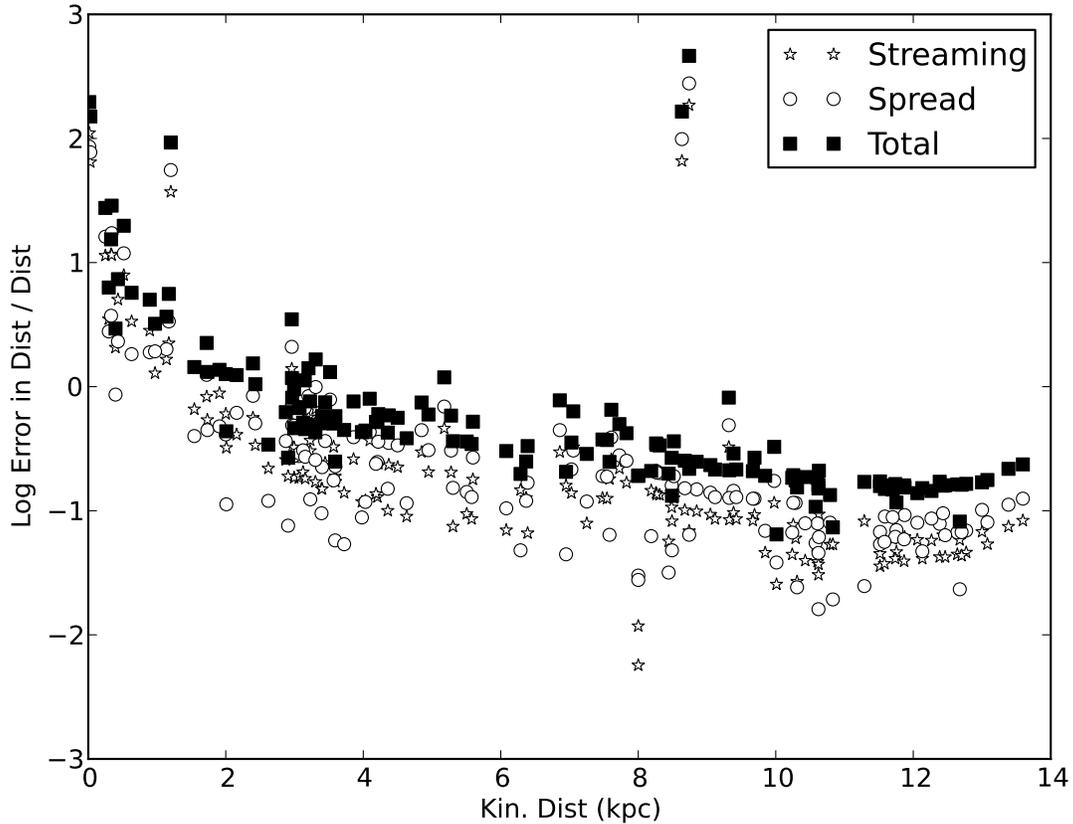}
\caption{\label{edist} Logarithm of fractional error in kinematic distance $\Delta D/D$ versus  kinematic distance $D$. Open stars correspond to error solely due to the streaming motion of ionized gas, open circles solely due to the expected velocity spread of $15\kms$, and filled squares the total error including the propagation of uncertainties in galactic longitude of the SFC $0.5$ deg, the uncertainties in the rotation curve, and the error due to spiral structure. The objects at $D\sim8\kpc$ with $\Delta D/D\sim100$ have forbidden values of $v_r$, so the errors in $v_r$ have been taken to be $\Delta v_r=\pm20\kms$. This plot excludes all the SFCs that were manually assigned either to $4\kpc$ or $12\kpc$.}
\end{figure}

It is also possible that the calculated LSR velocity of the SFC may be in error, affecting the kinematic distance measurement. Figure~\ref{edist} shows the error in distance $\Delta D$ as a function of kinematic distance $D$. The error contribution due to the (assumed) streaming motion of ionized gas is similar to the error due to the velocity spread inferred from (multiple) radio recombination line velocities. While there are a few cases where the fractional error is $\sim 10$ or even higher, the majority of the SFCs have a 36\% error in the kinematic distance. 

As a sanity check, the LSR velocity and the expansion velocity of SFC 264 in G353 can be compared to the results of~\citet{cappa11}, who find an expansion velocity $\sim 5-9~\kms$ which is in approximate agreement with the result quoted in this paper, $6.7~\kms$. Using a photometric distance to the cluster Pismis 24, which appears to power the HII region, they find $D=2.5$ kpc, substantially larger than our $D=0.9$ kpc. Photometric distances suffer from uncertainties in reddening, while kinematic distance estimates suffer from uncertainties in the velocity measurement and  the accuracy of the rotation curve. For example, the two distance estimates could be reconciled by correcting the LSR velocity $-3.8~\kms$ for a  $\sim10~\kms$ streaming motion of the ionized gas, yielding $v_r=-13.8~\kms$ and a kinematic distance measurement of 2.9 kpc. We conclude, therefore, that the two distance results agree within the uncertainties. We note that a study of CO map or any neutral gas tracer will provide a better estimate of both the expansion velocity and central velocity of the bubbles.

\subsubsection{Uncertainties due to Spiral Arms}
\label{sssec:spiral_arm}
\begin{figure}
\plotone{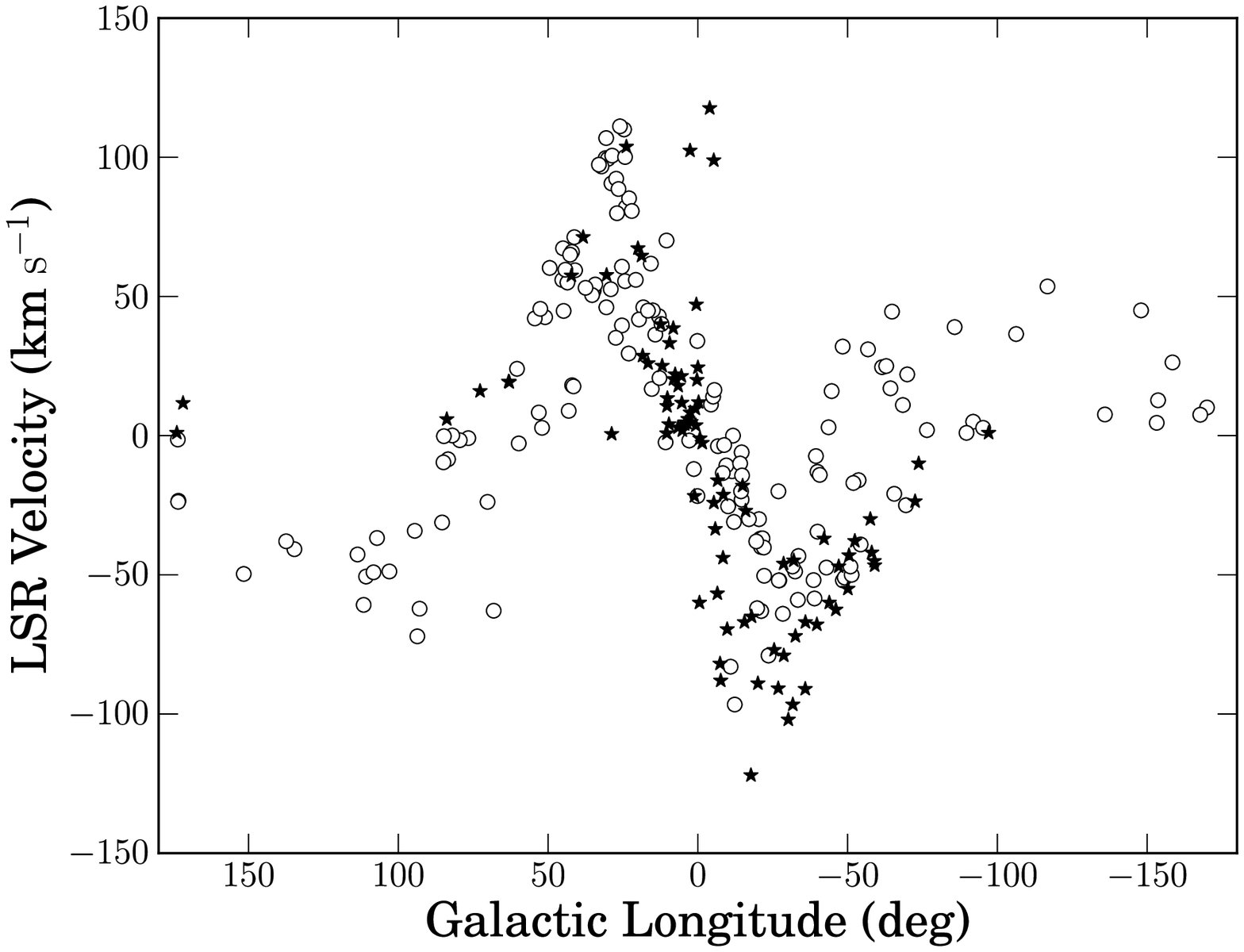}
\caption{\label{v_vs_l} LSR velocity plotted against galactic longitude of the centroid of the SFCs. Open circles denote the SFCs with kinematic distance ambiguity resolution while the filled stars denote those without resolution. All the SFCs with LSR velocity measurement are plotted. The slanted bar-like feature in $|l| \lesssim 30$ and a transverse velocity width of $\pm 20 \kms$ corresponds to the molecular ring~\citep[see][their figure 3]{dame01}. Star forming complexes in the ring account for $\sim50\%$ of our sources with $v_{LSR}$ measurements, $31\%$ of the Galactic free-free flux, and $35\%$ of the free-free luminosity of the Galaxy. This is likely an underestimate since, of the 127 SFCs in the ring, 59 SFCs do not have luminosity measurement due to the kinematic distance ambiguity. Note that the two sources with kinematic distance ambiguity at $l > 150^{o}$ have erroneous distances; they have forbidden radial velocities (negative velocity expected).}
\end{figure}

While the aforementioned derivation of kinematic distance assumes circular motion of the SFCs, the kinematics of Milky Way interstellar medium (ISM) are affected by the non-circular motions of spiral structures \citep{binney98}. \citet{gomez06} shows that the kinematic distances derived from a measured rotation curve have errors from $0.5\kpc$ to $3\kpc$ near the spiral arms. Taking $1.5\kpc$ as the typical error in the kinematic distance near the Galactic spiral arms~\citep[see][their Figure 5]{gomez06}, we add this error to the kinematic distance error propagation of SFCs that are likely associated with spiral arms.
Figure \ref{v_vs_l} shows the line of sight radial velocity versus the Galactic longitude of the SFCs. Comparing Figure \ref{v_vs_l} to Figure 3 of \citet{dame01}, we associate any SFCs that satisfy any of the following criteria with a spiral arm:

\begin{itemize}
\item Local Arm: $100^{o} \leqq l \leqq 160^{o}$ and $|v_{LSR}| \leqq 20\kms$
\item Perseus $\&$ Outer Arm: $30^{o} < l \leqq 180^{o}$ and $v_{LSR} < -20\kms$
\item Carina $\&$ Perseus Arm: $180^{o} \leqq l \leqq 330^{o}$ and $v_{LSR} > 20\kms$
\item Molecular Ring: $|l| \leqq 30^{o}$ and $|v_{LSR,rot}| \leqq 20\kms$
\end{itemize}

where $v_{LSR,rot}$ corresponds to $v_{LSR}$ rotated in the v-$l$ plane such that the molecular ring structure in Figure \ref{v_vs_l} is aligned along the $v_{LSR} = 0\kms$ axis. We include the molecular ring in the uncertainty analysis as well since it is likely composed of multiple spiral arms rooted from the Galactic central bar~\citep{dame01}.

\subsubsection{Catalogue of SFCs}
\label{sssec:sfccat}
The complete SFC catalogue is presented in Table \ref{sfclist} with the columns as follows: Column 1 gives the catalogue number, Column 2 the host WMAP source number, Columns 3 and 4 galactic longitude and latitude, Columns 5 and 6 semi-major and semi-minor axes, Column 7 position angle, Column 8 mean LSR velocity, Column 9 measured half-spread velocity, Column 10 estimated radial expansion velocity, Column 11 kinematic distance, Column 12 error in the kinematic distance, Column 13 effective radius (geometric mean of the semimajor and semiminor axis), Column 14 free-free flux, Column 15 dynamical time defined as the average bubble radius divided by the corrected expansion velocity, Column 16 the result of KDA resolution, and Column 17 the references used to resolve KDA.

The error in the kinematic distance is determined from the propagation of errors in the determination of Galactic longitude ($\Delta$ $l$ = $0.5^{o}$), rotation curve \citep[$\Delta$ $\sigma$ = 5$\kms$, ][]{clemens85}, 10$\kms$ streaming motion, spread in the velocity taken as measured half-velocity, corrected half-velocity if determined, or 15$\kms$ if the measured half-velocity was not determined or was less than 3$\kms$ (RM10 considers a velocity spread less than 3$\kms$ as multiple measurement of the same \ion{H}{2} region), and the uncertainties in the kinematic distance for SFCs associated with spiral arms or the molecular ring. In case of manual distance designation (for instance, some of the known SFCs near the Galactic centre; these SFCs are flagged in Table \ref{sfclist}) or for SFCs with no valid error calculation due to inaccurate $v_{LSR}$, the error in the distance is defined as 30$\%$ of the kinematic distance, following RM10. The median error contribution of each component is $1\%$ for galactic longitude, $2\%$ for rotation curve, 14$\%$ for streaming motion, 16$\%$ for the expansion velocity, and 21$\%$ for the spiral arm (only for the sources within 1kpc of the spiral arm). While spiral arm seems to be the most significant error contributor, only 72 of the 153 SFCs (all the SFCs with distances calculated by the rotation curve, not by manual assignment) were within 1 kpc of the spiral arm. We conclude that in the most general case, the streaming motion and the expansion velocity are the most significant error contributor. The overall median distance error is $35\%$.

The RRL velocity measurements of the individual \ion{H}{2} regions are presented in Table \ref{hiilist} where Column 1 gives the corresponding SFC number, Column 2 the name of the \ion{H}{2} region, Columns 3 and 4 the galactic coordinates, Column 5 the LSR velocity, and Column 6 the reference.

\subsection{Comparison to RM10}
\label{ssec:comp}
\begin{figure}
\plottwo{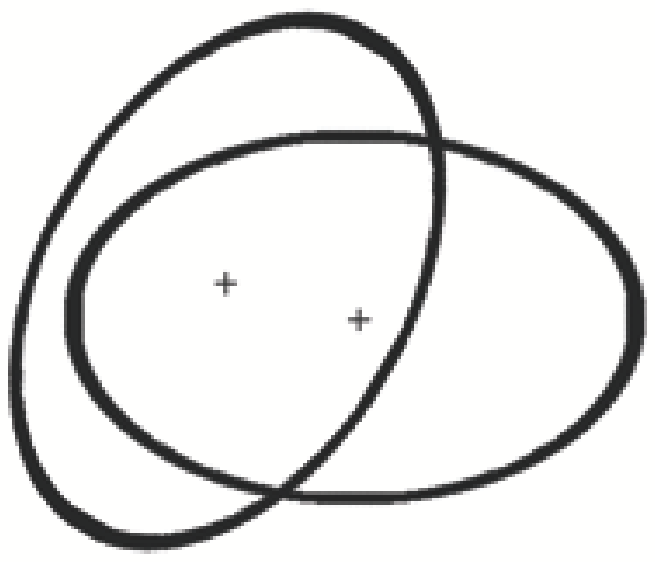}{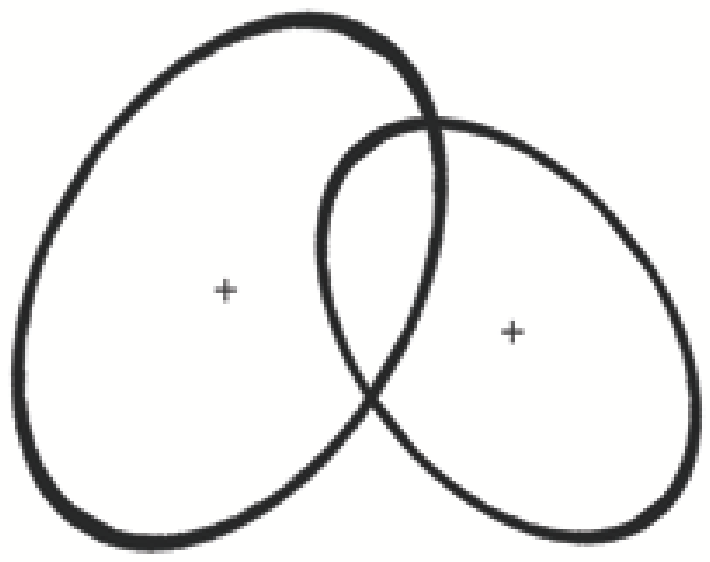}
\caption{\label{typs} Schematics of Type 1 and Type 2 matches between SFCs where crosses are the centroids of the SFCs. Left: Type 1. Right: Type 2.}
\end{figure}

All the SFCs, including those in the 13 WMAP sources that are discussed in RM10, are independently selected to ensure homogeneous classfication. As shown in Figure~\ref{wmapsrc}, by resizing the 13 WMAP sources studied by RM10, a total of 74 SFCs are identified compared to the previous 40: 19 original SFCs are kept, 18 are resized, reoriented, split or merged, and 33 are completely new. For a fair comparison between RM10 and this study, matches between the original and the new sample are divided into two groups: those that have centroids overlapping other SFCs (hereafter, Type 1), and those that only have an intersection between ellipses (hereafter, Type 2). Schematics showing these two different types are shown in Figure~\ref{typs}. As shown in Table \ref{sfccomp}, Type 1 matched objects have, on average, distance estimates that differ by 16$\%$. This corresponds to an error that is three times smaller than the typical error in kinematic distance in the SFC catalogue, 3.0$\kpc$. Using the average distance to our SFCs, the difference in ($l$,$b$) centroids between the two catalogs yields a difference of about 9pc in the bubble centroid, which is substantially smaller than the typical radius of massive GMCs\footnote{Since most stars are believed to be born in GMCs, it is likely that the SFCs presented here are embedded in GMCs. \citet{gmc}, for instance, matches the SFCs to known GMCs to estimate the star formations efficiencies and lifetimes of the GMCs.}. Comparing the semi-major and semi-minor axes between our sample and RM10 shows that the SFCs as identified in this paper are somewhat larger. The two samples identify either the same or physically associated objects, where some of RM10 SFCs are inside the SFCs found in this work. However, we note that there is a non-negligible fraction of un-matched complexes since we find 33 new SFCs out of 74. Type 2 matches, in general, are significantly more discrepant than Type 1 matches, but it is likely that the Type 2 matches are physically different objects.

\section{Results}
\label{sec:results}

\subsection{SFC Free-Free Luminosity}
\label{ssec:fflum}

Free-free emission is produced by photoionized gas both inside and outside the SFC shell walls. Assuming the origin of the ionization is associated with stars at the distance of the SFC, the total free-free luminosity of all the SFCs has a direct relationship to the galactic star formation rate (SFR). The energy per unit volume per unit time emitted by free-free emission in an ionized region is given by~\citep{karzas61}:

\begin{equation}
\label{eq:ffem}
\epsilon^{ff}_{\nu} = \frac{2^{5} \pi e^{6}}{3 m_{e} c^{3}}
\left(\frac{2\pi}{3\kb m_{e}}\right)^{1/2}
T^{-1/2}Z^{2}n_{e}n_{i}e^{-h \nu /\kb T}g_{ff},
\end{equation}
where Z is the charge per ion, $\kb$ is Boltzmann's constant, c the speed of light, $m_{e}$ the electron mass, T the electron temperature, $n_{e}$ and $n_{i}$ electron and ion densities, respectively, and $g_{ff}$ is the Gaunt factor. Following the procedure of MR10, we adopt $n_{e}$ = $n_{i}$, Z = 1, $T_{e}$ = 7000 K, typical values of \ion{H}{2} regions, and $g_{ff}$ = 3.28~\citep{sutherland} using 94GHz corresponding to W band WMAP free-free map. This corresponds to $\epsilon^{ff}_{\nu} = \epsilon_{0} n_{e}^{2}$ where $\epsilon_{0} = 2.67~\rm{x}~10^{-39} g~cm^{5}~s^{-3}~Hz^{-1}$.

The total number of ionizing photons required to keep an isotropic \ion{H}{2} region ionized is

\begin{equation}
\label{eq:qtoteq}
Q_{tot} = \int{n_{e}^{2}\alpha(H^{+})dV},
\end{equation}

where $\alpha(H^{+}) = 3.57~\rm{x}~10^{-13}\cm^{-3}\s^{-1}$ is the H recombination coefficient~\citep{osterbrock89} and V is the volume of the ionized region. The total ionizing luminosity of a given \ion{H}{2} region is then

\begin{equation}
\label{eq:qtot_lv}
Q_{tot} = \frac{\alpha(H^{+})}{\epsilon_{0}} L_{\nu} \approx 1.34~\rm{x}~10^{26}~\left({L_{\nu}\over {\rm erg\, s}^{-1}}\right)~\rm{photons}~s^{-1}.
\end{equation}

where $L_{\nu}$ is the specific free-free luminosity (ergs per second per Hz),  at 94GHz, of the given \ion{H}{2} region.

\subsubsection{Galactic Distribution of SFC}
\label{sssec:sfcdistr}

\begin{figure}
\plottwo{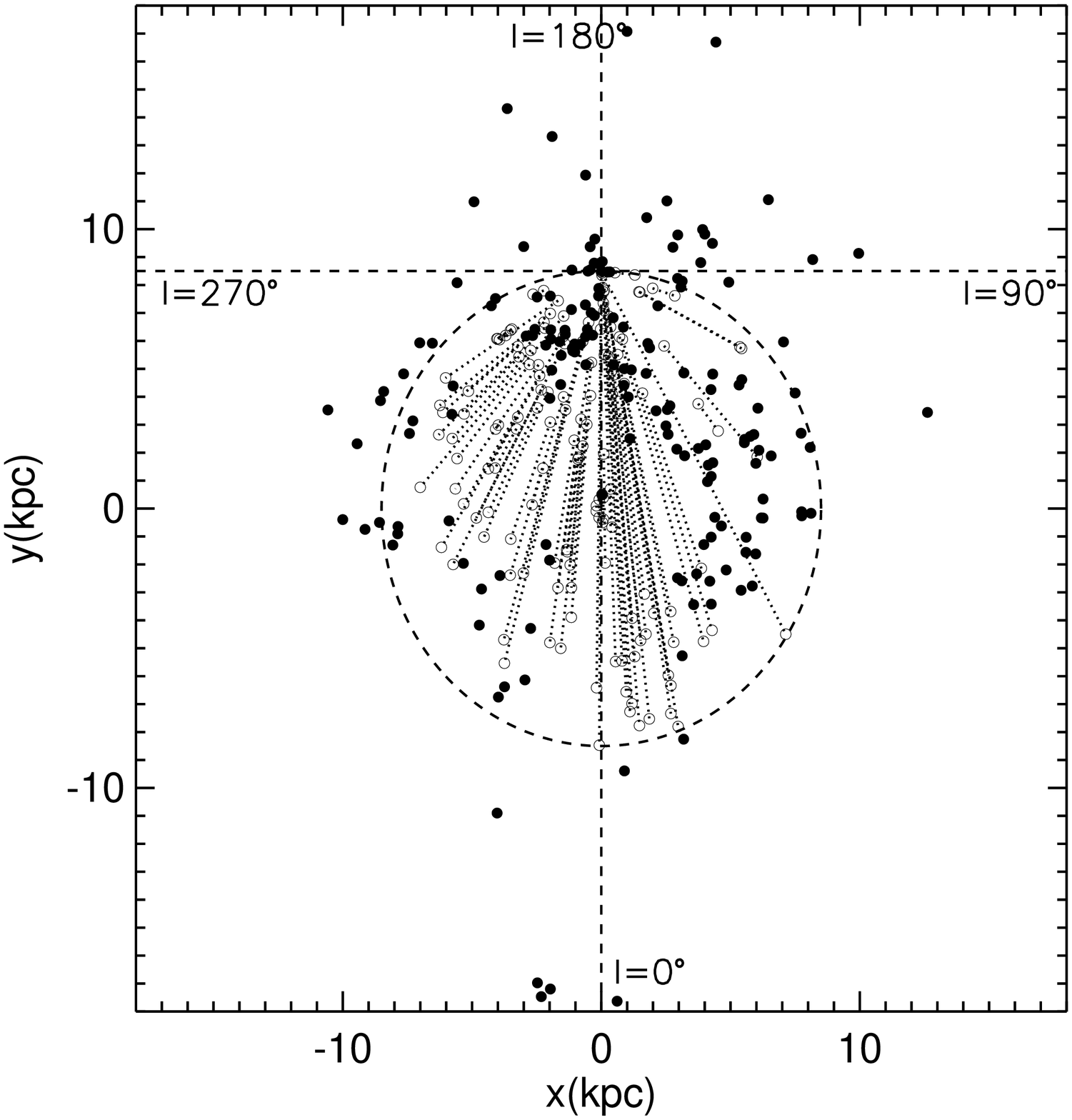}{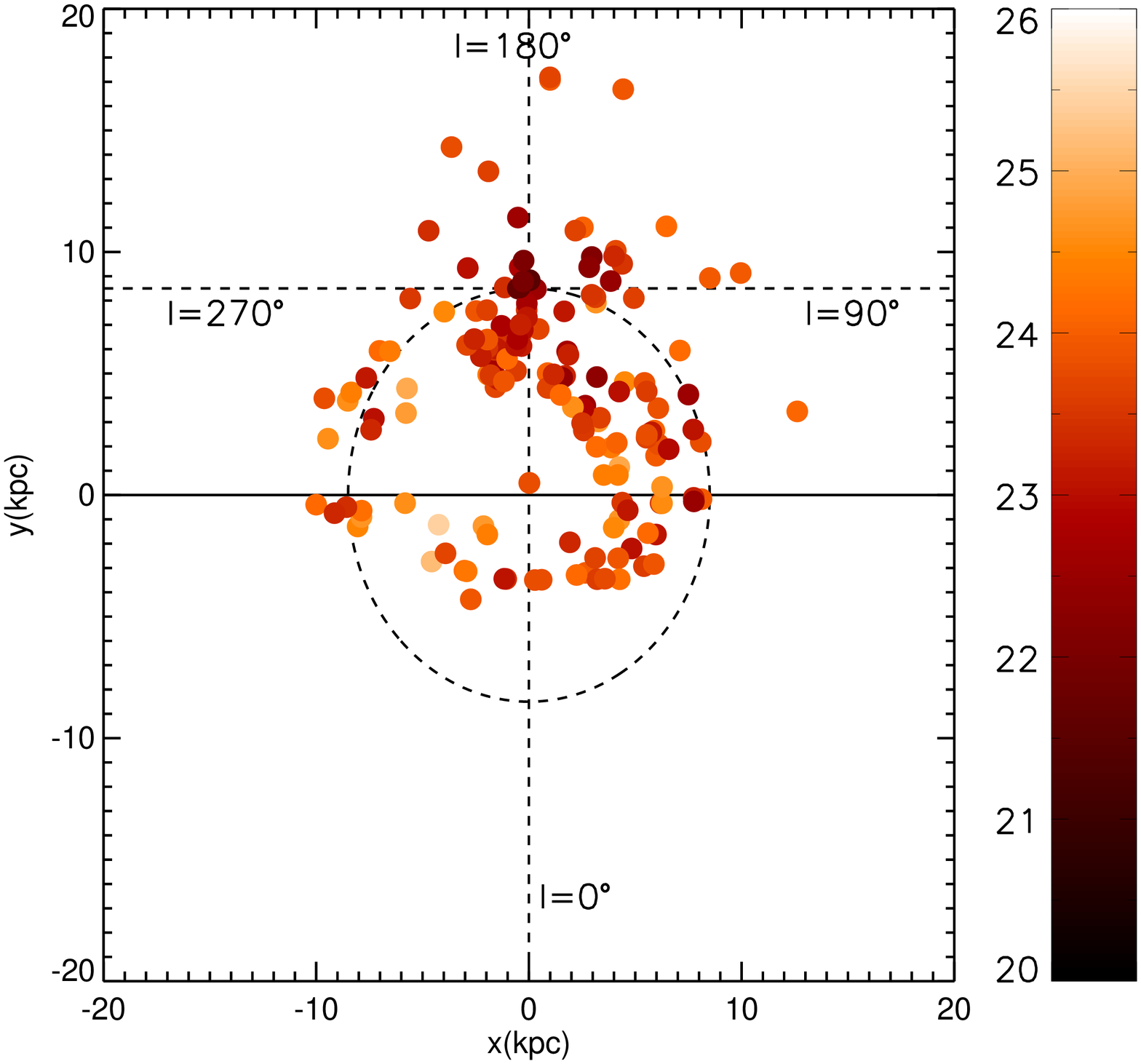}
\caption{\label{sfcdistr} Left: Galactic distribution of all the SFCs with kinematic distances. The dashed circle is the solar circle and its intersection with the two dashed lines is the position of the sun. Near and far distances of a single source are connected with dotted lines. A solid circle denotes SFCs with a unique distance measurement or KDA resolution. Right: Galactic distribution of all the SFCs with free-free luminosity determination. Colours represent the free-free luminosity in $\rm{erg}~\rm{s}^{-1}~\rm{Hz}^{-1}$ on a logarithmic scale.}
\end{figure}

The galactic distribution of all the SFCs in the catalogue (left panel) and only those with unique distance and free-free flux measurement (right panel) are shown in Figure \ref{sfcdistr}. We note that the SFC distribution in Figure \ref{sfcdistr} shows non-axisymmetric features. This is partly due to the paucity of RRL observations in the fourth quadrant compared to the first quadrant. It is also possible that the SFCs are tracing the non-axisymmetric matter distribution in the Milky Way galaxy, previously found by \citet{dame87}. 

Past studies have shown that most star formation in the Galaxy occurs in molecular clouds, and the Galactic distribution of molecular hydrogen falls off rapidly beyond the solar circle~\citep{dame01, binney98}. Thus we should expect that the star formation falls off with distance from the galactic center, at least beyond the molecular ring. This behavior is apparent in Figure \ref{v_vs_l} where $\sim 50\%$ of all the SFCs with LSR velocity measurement, and $35\%$ of the free-free luminosity, are associated with the molecular ring, which is at Galacto-centric radius $\sim 3.6 \kpc $~\citep{binney98}. As expected, Figure \ref{sfcdistr} shows that there are significantly fewer SFCs outside the solar circle (hereafter, OSC) than those inside the solar circle (hereafter, SC). 

We have compiled the luminosity function of star forming complexes inside and outside the solar circle in Figure \ref{lum_fn}. A Kolmogorov-Smirnov (KS) test shows that the normalized luminosity functions do not differ at $\sim$ 100$\%$. 

While the shapes of the luminosity functions inside and outside the solar circle are similar, the total free-free luminosity of sources inside the solar circle is $1.0~\rm{x}~10^{27}$ $\rm{erg}~\rm{s}^{-1}~\rm{Hz}^{-1}$; that of sources outside the solar circle is only $0.2~\rm{x}~10^{27}$ $\rm{erg}~\rm{s}^{-1}~\rm{Hz}^{-1}$. Furthermore, the distances to sources outside the solar circle may well be overestimated by a factor of $1.5$, as mentioned in $\S$ \ref{ssec:prop_sfc}, so that the corrected OSC cumulative luminosity might be as little as $\sim$ 9$\%$ of the cumulative luminosity of sources inside the solar circle.

\subsubsection{A Possible Distance Bias}
\label{sssec:bias}

\begin{figure}
\plotone{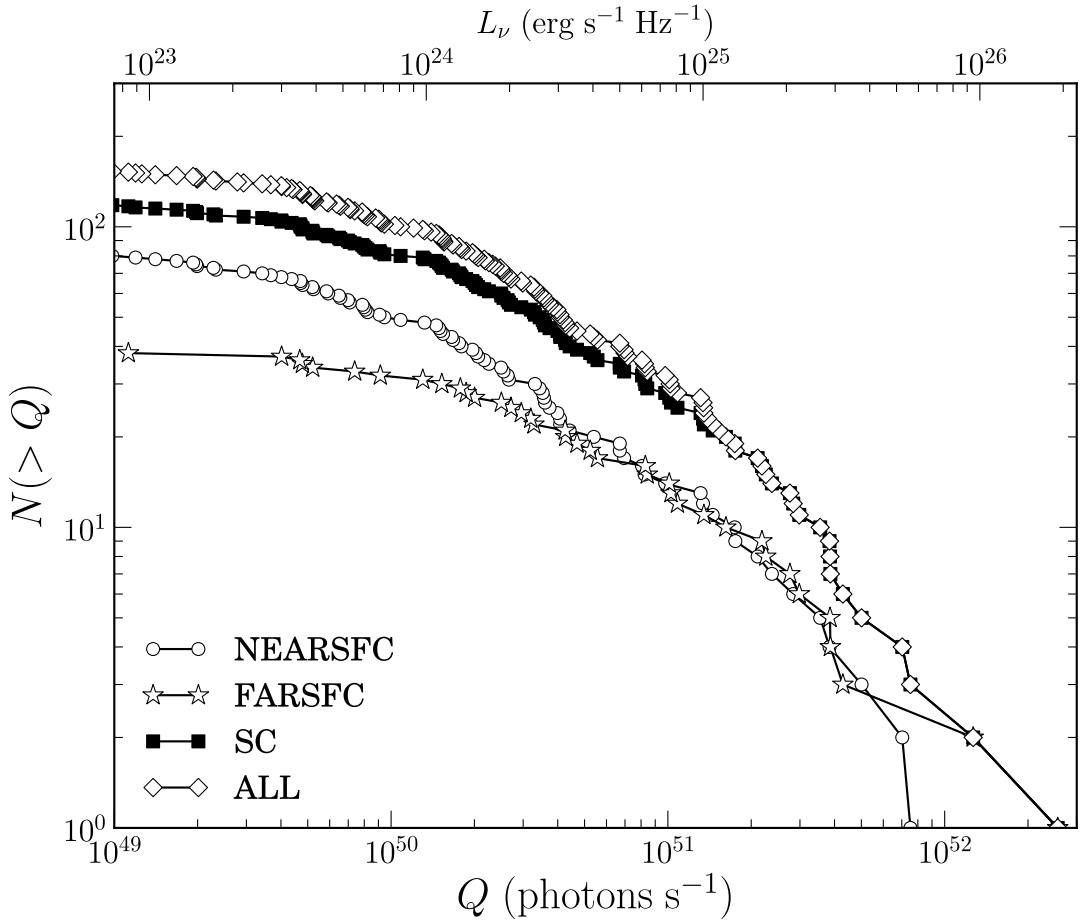}
\caption{\label{lum_fn} The cumulative ionizing luminosity function (dust-uncorrected) of SFCs with unique distance and flux measurement. Open circles denote SFCs inside the solar circle on our side of the Galaxy ($y>0$ in figure \ref{sfcdistr}; labeled NEARSFC): solar circle SFCs on the far side (y $\leqq$ 0) are denoted by open stars (FARSFC). Filled squares denote all objects inside the solar circle (SC). The open diamonds (ALL) denote all the SFCs, both inside and outside the solar circle, with the distance correction factor of 1.5 as discussed in $\S$ \ref{sssec:vel_dist} applied when calculating $Q$.}
\end{figure}

\begin{figure}
\plotone{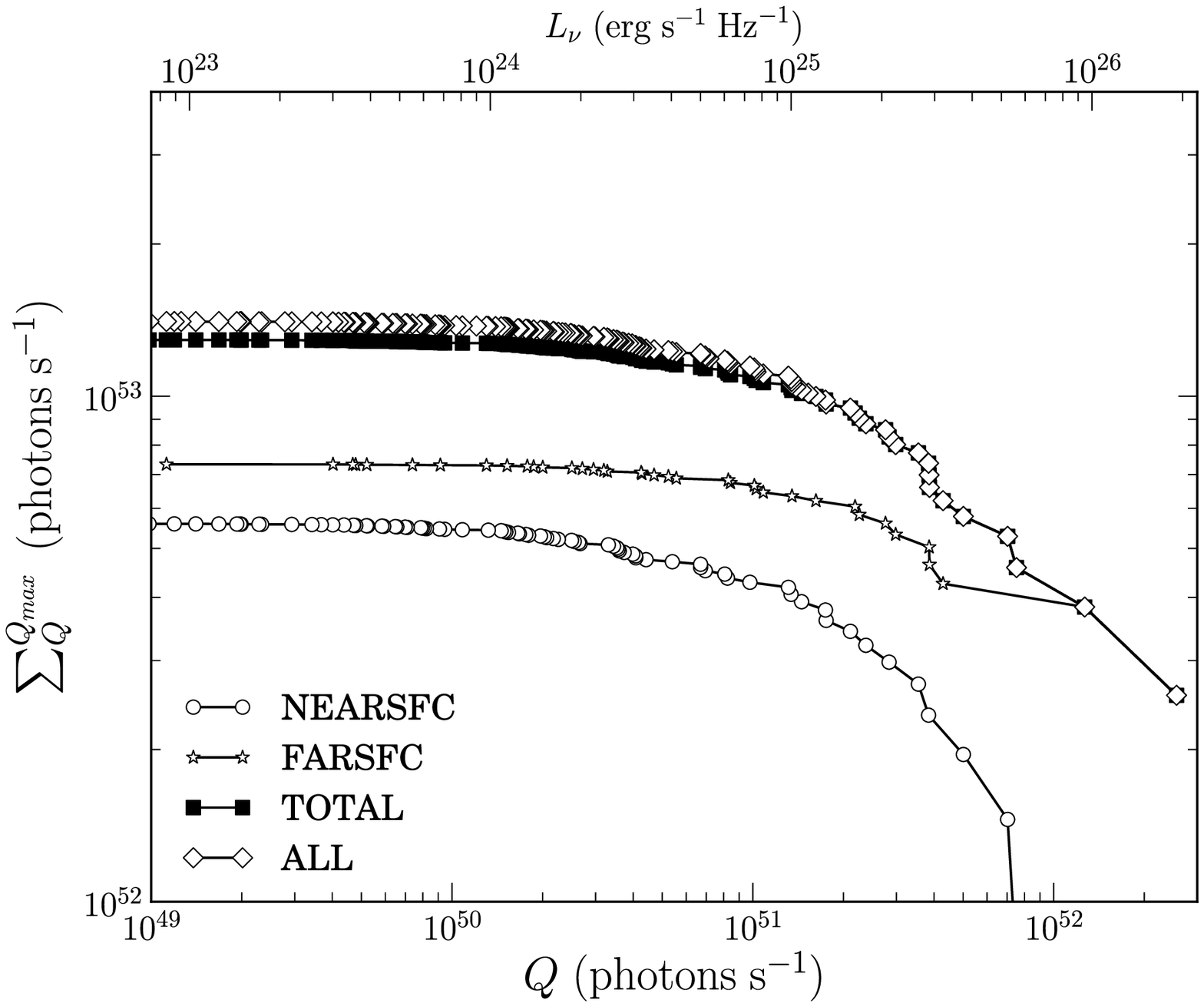}
\caption{\label{lum_fn_sum} Cumulative ionizing luminosity versus ionizing luminosity (dust-uncorrected). The legend is the same as Figure \ref{lum_fn}. Note that the cumulative luminosity of sources inside the solar circle, but on the far side of the galaxy from the sun, exceeds that of sources in side the solar circle but on the near side of the galaxy by $\sim31\%$. The 24 most luminous objects (those with $Q>1.3\times10^{51}\s^{-1}$) produce a total ${\cal Q}=1.5\times10^{53}\s^{-1}$, half the (non-dust corrected) ionizing flux of the entire Galaxy.}
\end{figure}

Figure \ref{sfcdistr} shows that the surface density of star forming complexes is higher near the sun than at large heliocentric distances, suggesting that we may be missing sources at large distances and so underestimating the Galactic free-free luminosity. To see if this is the case, we examine the luminosity function and the cumulative luminosity of sources on the near and far side  of the Galaxy. By the near side of the Galaxy we mean sources with $y\geq 0$ in Figure \ref{sfcdistr}; sources with $y<0$ are then on the far side of the Galaxy. The luminosity function and cumulative luminosity as a function of $Q$ (for near-side star forming complexes, far-side complexes, all complexes inside the solar circle, and all the SFCs) are compared in Figure \ref{lum_fn} and Figure \ref{lum_fn_sum}, respectively. 

Figure \ref{lum_fn} shows that the total number of far-side sources is 38, less than half the total number of those on the near-side (84), consistent with the impression given by figure \ref{sfcdistr}. Furthermore, Figure \ref{lum_fn} shows a turnover at $Q \sim 10^{50}\s^{-1}$---the break is seen more clearly in the differential luminosity function, to be discussed below. In fact, we find no sources with $Q<10^{49}\s^{-1}$ on the far side of the Galaxy. This demonstrates that we have a selection bias against low luminosity ($Q<10^{49}\s^{-1}$) sources at large distances, a result similar to that found by \citet{paper1}.

Despite this, Figure \ref{lum_fn_sum} shows that the total ionizing luminosity of sources on the far-side is $\sim 31\%$ {\em greater} than that of near-side sources. In fact, Figure \ref{lum_fn_sum} shows that the 24 most luminous sources (with dust-uncorrected $Q\gtrsim1.3\times10^{51}\s^{-1}$), out of 280 total sources ($\sim9\%$) account for a cumulative (dust-uncorrected) ${\cal Q}\approx1.5\times10^{53}\s^{-1}$, or $75\%$ of the luminosity from sources with known distances. These sources account for half of the total free-free luminosity of the Galaxy, as determined below. So while it is true that our selection has a bias against distant, low luminosity sources, low luminosity sources make an insignificant contribution to the total free-free emission of the galaxy. 

\begin{figure}
\plotone{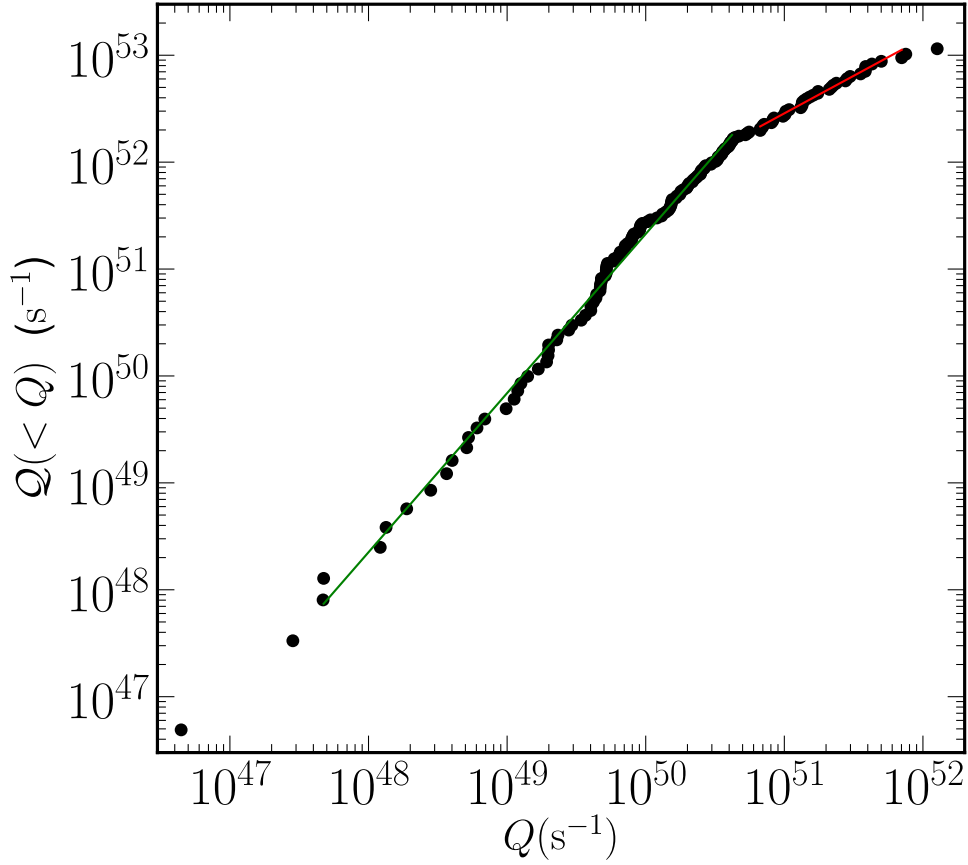}
\caption{\label{lum_fn_sum_reversed} The cumulative ionizing luminosity ${\cal Q}(<Q)$ versus ionizing luminosity $Q$. The 24 most luminous objects (those with $Q>1.3\times10^{51}\s^{-1}$) produce a total ${\cal Q}=1.5\times10^{53}\s^{-1}$, half the (non-dust corrected) ionizing flux of the entire Galaxy.}
\end{figure}

Figure \ref{lum_fn_sum_reversed} shows the same data as in figure \ref{lum_fn_sum}, but this time integrating from the minimum $Q$ to large $Q$, making it easier to measure the slope of the cumulative luminosity function
\be  
{\cal Q}(<Q)\equiv\int_0^QQ{dN\over dQ}dQ.
\ee  
The data can be described by two power laws, one for $Q>6.7\times10^{50}\s^{-1}$, and another for $Q<4.2\times10^{50}\s^{-1}$ (there is a hint of a third break at $Q<10^{50}\rm{s}^{-1}$ but it is not very clear). Following \citet{mckee} and \citet{paper1}, we define the power law exponent $\Gamma$ by
\be
Q{dN\over dQ}\propto Q^{-\Gamma},
\ee
and ${\cal Q}\propto Q^{\tilde\Gamma}$, where $\tilde\Gamma=1-\Gamma$. The red line in figure \ref{lum_fn_sum_reversed} is a least squares fit to the data above $Q\approx6.7\times10^{51}\s^{-1}$, with $\tilde\Gamma\approx0.7$. The green line shows that $\tilde\Gamma\approx1.5$ for $Q\lesssim4.2\times10^{50}\s^{-1}$. In other words, for $Q<4.2\times10{50}\s^{-1}$, $\Gamma=-0.5$, while for larger $Q$ we find $\Gamma=0.3$.

As a check, figure \ref{mc_data_dndl_all} shows the differential luminosity function $dN/dQ$. Again, we find two different slopes, defined as $(\Delta \log dN)/(\Delta \log Q)=-\Gamma$. for $Q<4.2\times10^{51}\s^{-1}$, we find $\Gamma\approx-0.5$ and $\Gamma\approx0.7$ for larger values of $Q$. These exponents are in reasonable agreement with the corresponding values obtained from figure \ref{lum_fn_sum_reversed}. While $\Gamma$ of smaller $Q$'s are in perfect agreement, the exponents for larger values of $Q$ agree within $\sim\pm0.4$ precision. We note that half of the bins in the high $Q$ regime in \ref{mc_data_dndl_all} are on the order of 1 or 2 and suffer from poor sampling. In fact, we find the exponents to change by a factor of two within two decades of binsizes. We argue that the two luminosity functions are in agreement within the precision of the sampling.

The slope $\Gamma=0.3-0.7$ of the high luminosity tail is similar to the values, ranging from $0.41$ to $1.3$, with an average $\Gamma=0.75$, found in other galaxies by \citet{KEH} and \citet{mckee}. While the slope we find is shallower than the mean value found by \citet{delgado97}, it is in agreement with their slope for later type galaxies \citep[see][their Figure 3]{delgado97}. The value of $\Gamma$ we find for the Milky Way is consistent with our result that $24$ sources are responsible for half the ionizing luminosity of the Galaxy.

We conclude that our sample is biased against finding low luminosity sources at large heliocentric distances (larger than a few $\kpc$), but that this bias has only a minor effect on our estimate of the Galactic free-free or ionizing luminosity.

\subsubsection{Accounting for sources with unknown distances}

The 168 SFCs that have both unique distance and free-free flux measurement only account for 58$\%$ of the total Galactic free-free flux. The 280 SFCs that have free-free flux measurements (but not necessarily unique distance measurements) account for 74$\%$ of the total Galactic free-free flux (see \S \ref{sssec:ff-fluxes} above). To estimate the luminosities of the 112 SFCs with known flux but unknown distances, we perform a Monte Carlo (MC) simulation. To do so, we need a model for the galactocentric radius distribution. We assume that the star formation is distributed in a manner similar to that found for molecular gas, as shown, for example, in Figure 9.19 of \citet{binney98}. In particular, we model the SFC distribution as a Gaussian with mean radius  $\mu = 5\kpc$ and dispersion  $\sigma = 5.7\kpc$. These parameters, and the dispersion in particular, are found by requiring that the number of SFCs inside and outside the solar circle match the observed numbers for our sample of 168 sources. For each SFC without distance, a galactocentric radius is randomly sampled from the gaussian distribution reflecting the variations of the molecular gas. If the MC-generated radius is not allowed at the measured Galactic longitude of the SFC, we repeat the process until we find an allowed radius.

Fixing the Galactic radius and longitude of the MC source leaves us with a Heliocentric distance ambiguity if the source is inside the solar circle. To resolve this ambiguity, we apply three criteria. First, the free-free luminosity of the SFC must be greater than the minimum luminosity detectable at that distance. We estimate this minimum luminosity from the data; Figure \ref{dbias_minl} shows the minimum detected luminosity as a function of Heliocentric distance for the 166 sources with known distances. We fit a power law to the data, and use this fit to determine the minimum detectable luminosity as a function of distance. We find that this minimum detectable luminosity scales roughly as $D^2$ (the actual value of the exponent is $1.96$) suggesting that there is in fact a roughly fixed {\em flux} limit. We refer to the requirement that our MC sources exceed this minimum luminosity as the sensitivity check.

The second criterion we use to resolve the distance ambiguity relies on our knowledge of Galactic structure; we note that SFCs have a higher chance of being associated with spiral arms than of being between arms. If the distance ambiguity is not resolved by the sensitivity check, then the distance that is within $1\kpc$ of an arm in the galactic spiral arm model of \citet{taylor93} is chosen; we call this the spiral arm check. 

Finally, we expect most, if not all, SFCs to lie within the Galactic plane. We assert that the height of the SFC above the plane, z, calculated using near distance ($z_{near}$) must be smaller than 60pc (scale height of the Galactic molecular disk), and the z calculated using far distance ($z_{far}$) be smaller than 200pc. If the ambiguity remains after sensitivity and spiral arm check, and $z_{near} \leq 60\rm{pc}$ and $z_{far} > 200\rm{pc}$, we adopt the near distance; we call this height check. These parameters are found by requiring that the number of SFCs on our side and the other side of the Galaxy inside the solar circle match the observed number for our sample of 168 sources. While we expect most SFCs to lie within the scale height of the molecular disk, 60pc, above the plane, especially for those SFCs associated with the molecular ring, we leave the possibility of outliers. In fact, 16 SFCs out of the 168 SFCs with unique distance measurement have z $>$ 60 pc. Of these 16 SFCs, 6 of them are likely associated with the molecular ring, and only one of these 6 SFCs is at z $>$ 100 pc (but its Galactic latitude is $1.8^{o}$). We also note that the height has a mean uncertainty of $\sim35\%$ due to the uncertainties in the distance derived from RRL velocities.

If both distances pass or fail the height check, we choose the distance assuming a uniform probability distribution between the two distances, as we have no other reason to favour one side over the other. 

\begin{figure}
\plotone{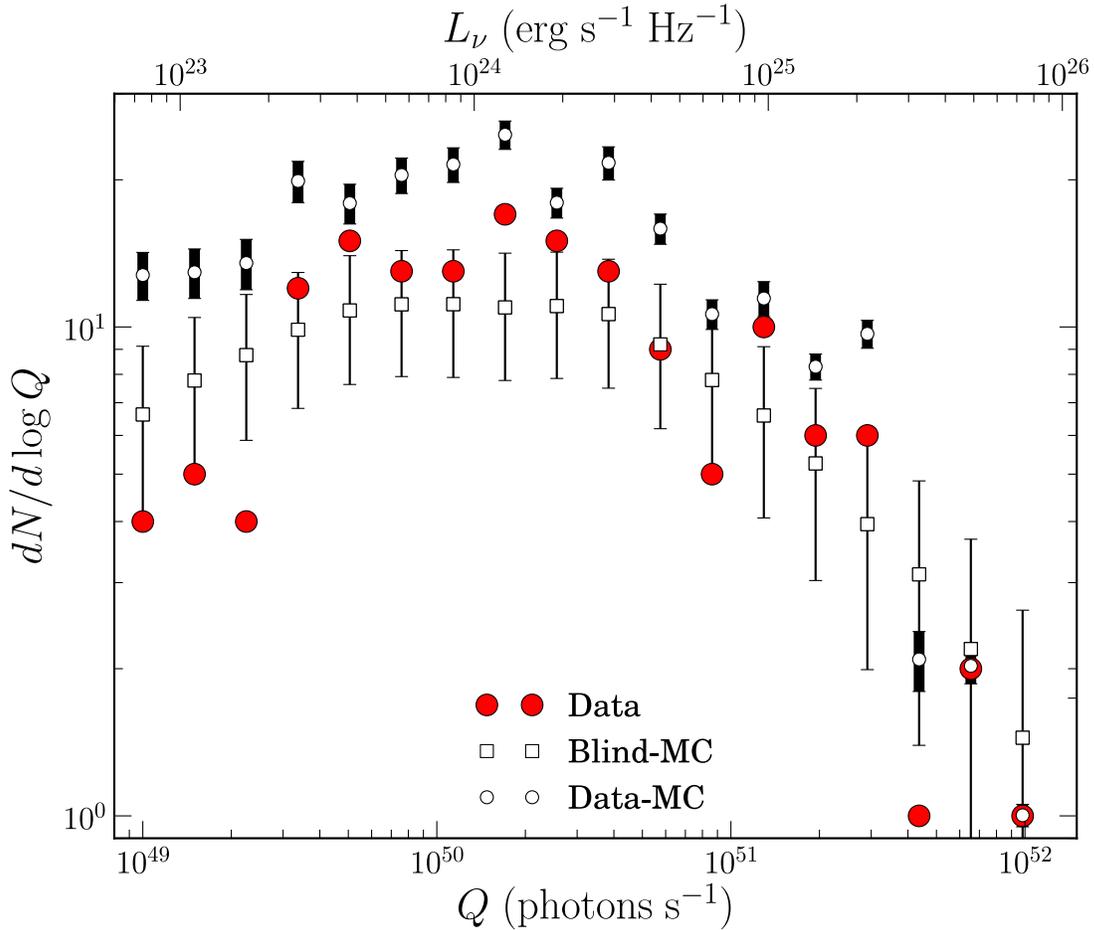}
\caption{\label{mc_data_dndl_all} The differential luminosity function of SFCs . The y-axis gives the number of SFCs in $\log Q = 0.176$ bins. Filled (red) circles are the observed SFC sample (166 objects) while open squares represent the MC simulated SFCs. A KS test shows that the two samples do not disagree at the 20$\%$ confidence level. The open circles with  error bars represent the differential luminosity function generated by combining the 168 objects with known luminosities with a MC model for the 112 objects lacking measured distances. 
}
\end{figure}

\begin{figure}
\plotone{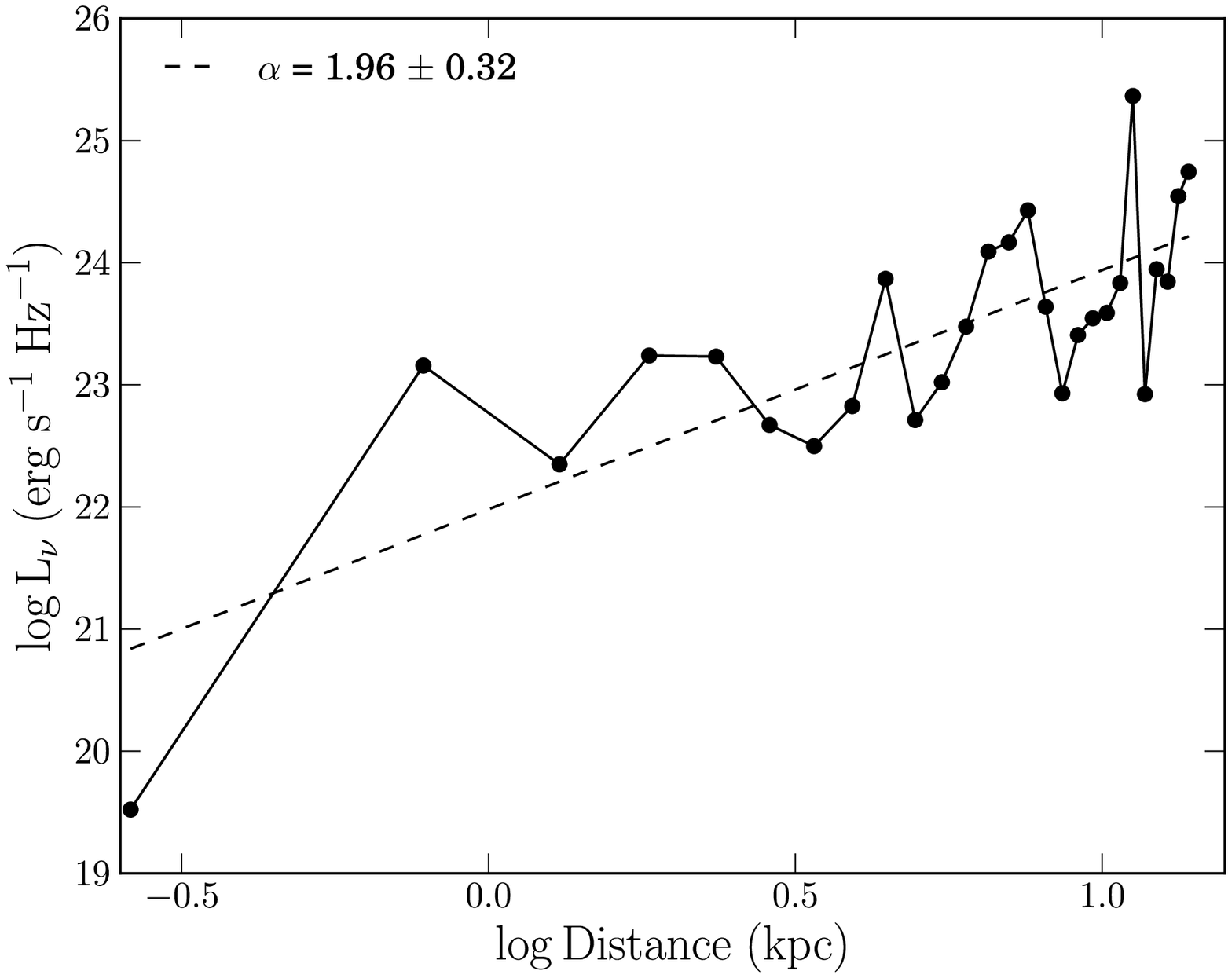}
\caption{\label{dbias_minl} The minimum luminosity of observed SFCs plotted against the (binned) kinematic distance. The slope of the fitted line is $\alpha\approx1.96$.}
\end{figure}

To check our model, we perform MC simulations with 168 SFCs having galactocentric radii drawn from the normal distribution described above. We then assign a galactic longitude chosen randomly from the allowed range governed by the galactocentric radius. Galactic latitudes are drawn from the normal distribution with $\mu = 0.11^{o}$ and $\sigma = 0.90^{o}$ where the parameters are the results of the gaussian fit of the full sample of 280 SFCs. Finally, we assign a free-free flux randomly from the catalogue of fluxes from the 168 SFCs with distances. Near and far distances of the SFCs within the solar circle are then assigned based on the three checks described above.

Repeating this process 5000 times result in, on average, 91 SFCs are on the near side of the Galaxy, and 30 on the far side, all inside the solar circle. This leaves 47 SFCs outside the solar circle. This is similar to the observed numbers, by construction. A KS test shows that  the simulated and observed differential luminosity functions do not disagree at the $20\%$ confidence level. Figure \ref{mc_data_dndl_all} shows the differential luminosity function of the 168 observed SFCs with known luminosities (filled circles), as well as that generated by our MC model (open squares). 

Repeating the MC simulation 5000 times for the remaining 112 SFCs which have free-free flux measurements but lack distance determinations, the model predicts 94 SFCs inside the solar circle, with 80 on our side of the Galaxy and 14 on the far side. 
%
%
The free-free luminosity of the Galaxy (``All''), of sources inside the solar circle (``SC'') and on the near and far side (but inside the solar circle) predicted by the MC simulations, are all summarized in Table \ref{sfcsimres}, which also lists the observed luminosities. The table heading ``Blind-MC'' refers to the MC sample of 168 SFCs with randomly assigned Galactocentric radii (chosen from the normal distribution described above), longitudes chosen randomly from the allowed range, latitudes chosen from the normal distribution described above, and free-free fluxes randomly assigned from the measured fluxes of the 166 sources. The table heading ``Data-MC'' refers to the combination of the 166 SFCs with measured luminosities, supplemented by the MC simulation of 112 SFCs using the observed Galactic longitudes and free-free fluxes.

The MC estimate of the free-free luminosity (Blind-MC) agrees with the observed free-free luminosity (both for 168 SFCs) at the one sigma level, as shown in Table \ref{sfcsimres}. We conclude that our MC model is reliable at the 1$\sigma$ level, and henceforth use the luminosities from the third column of table \ref{sfcsimres} (for all 280 SFCs) as our best estimate for the free-free luminosities from the corresponding regions of the Galaxy.

\subsection{Luminosity and SFR}
\label{sssec:lum_SFR}
The Galactic free-free luminosity due to star forming complexes is given by the sum of the free-free luminosities of the 168 sources with measured fluxes and distances, supplemented by the Monte Carlo estimates for the luminosities of the 114 SFCs with measured fluxes but not distances. We find
\begin{equation}
\label{eq:lumsfc}
L_{\nu, SFC} = 1.3 \pm 0.2~\rm{x}~10^{27}~erg~s^{-1}~Hz^{-1}
\end{equation}
at W band, where the error estimate includes the error in the kinematic distance as described in $\S$ \ref{sssec:sfccat} added in quadrature to the 1$\sigma$ error of the Monte Carlo simulation. The luminosity of objects outside the solar circle are reduced by the factor of $1.5$ described in $\S$ \ref{sssec:vel_dist}; without this correction, $L_{\nu, SFC}$ is larger than that given in eqn. \ref{eq:lumsfc} by $\sim$ 10$\%$. 

We note that our estimate of the ionizing luminosity of the SFC associated with the Arches cluster (G359 SFC0) is $1.8\times10^{51}\s^{-1}$ which is about 4 times less than expected assuming Arches cluster has 150 O3 main sequence stars ($6.5\times10^{51}\s^{-1}$). This is likely due to an underestimate of the free-free flux assigned to SFC0. This is a warning that, as the Galactic Centre is a highly confused region, the SFC parameters in that region should be taken with caution.

The total free-free flux in 280 SFCs is $\rm{F}_{\nu, SFC}$ = 34234.8 Jy, or 74$\%$ of the galactic free-free flux $\rm{F}_{\nu, gal}$ = 46177.6 Jy. This result is consistent with Figure \ref{wmapsrc}, which shows that most of the free-free flux is concentrated along the Galactic plane where we find SFCs. We believe that the remaining 26$\%$ of the flux arises from the diffuse component of the free-free emission leaking out of the WMAP sources. If so, then the unaccounted for flux is emitted, on average, at very nearly the same distance as that emitted by the SFCs. Hence we estimate the total Galactic free-free luminosity (due to both SFCs and to diffuse emission) to be
\begin{equation}
\label{eq:lumgal}
L_{\nu} = L_{\nu, SFC} + 4 \pi D_f^{2} (F_{\nu, gal} - F_{\nu,
  SFC})=1.6 \pm 0.3~\rm{x}~10^{27}~erg~s^{-1}~Hz^{-1},
\end{equation}
where
\be  
D_f \equiv {\sum_i D_i f_i\over \sum_i f_i} = 4.7 \pm 1.8 \kpc
\ee  
is the flux weighted mean distance to the $168$ SFCs for which we have both unique distance and flux estimates, supplemented by the Monte Carlo estimates for the luminosities of the 114 SFCs with measured fluxes but not distances (the value changes to $4.8 \kpc$ without the Monte Carlo estimates). If we assume the distance is $ 8\kpc$, as expected if most SFCs are in the $3 \kpc$ molecular ring), $L_{\nu}$ changes to $2.2 \pm 0.3\times10^{27}~\rm{erg}~\rm{s}^{-1}~\rm{Hz}^{-1}$, $38\%$ higher than the quoted value. Without the correction for the luminosity of the objects outside the solar circle, $L_{\nu}$ increases by $\sim8\%$.

The error estimate in equation \ref{eq:lumgal} comes from propagating the errors in the  distance estimates. The error in the mean distance contributes an error of $25\%$, the error in the distance of individual SFCs contributes another $8\%$, while the 1$\sigma$ error in the MC simulation accounts for $1\%$. In $\S$ \ref{sssec:sfccat} we showed that the most significant error in the kinematic distances were due to the expansion velocity of SFCs. We conclude that the most significant source of uncertainty in our estimate of the Galactic free-free luminosity is the expansion velocity of SFCs, combined with the uncertainty in the mean distance of the sources responsible for the diffuse flux.

Using Equation \ref{eq:qtot_lv}, the total ionizing luminosity, before the correction for dust absorption, is ${\cal Q}$ = 2.1~$\pm$~0.4~x~$10^{53}$~\rm{photons}~$s^{-1}$. Following \citet{mckee}, we correct for the effect of ionizing photons' absorption by dust grains by multiplying by 1.37:
\begin{equation}
\label{eq:qtotgal}
{\cal Q} = 3.0 \pm 0.5~\rm{x}~10^{53}~\rm{photons}~s^{-1}
\end{equation}
compared to the estimate of MR10, ${\cal Q}=3.2\pm0.5\times10^{53}\s^{-1}$. To calculate the star formation rate of the Milky Way, we use the conversion from ionizing flux to stellar mass given by MR10:
\be 
\dot M_* /Q = 4.1\times10^{-54}M_\odot\yr^{-1}\s.
\ee 
The star formation rate of the Milky Way is then
\begin{equation}
\label{eq:sfr}
\dot{M}_{*}  = 1.2 \pm 0.2~M_{\sun}~\rm{yr}^{-1},
\end{equation}
which differs from the estimate of MR10 by $\sim 8\%$. As MR10 note, however, the estimate given above is highly dependent on the choice of the IMF, in particular the high mass end slope (denoted by $\Gamma$). We follow MR10 and assume $\Gamma = 1.35$, the \citet{salpeter55} value.

This estimate for the star formation rate also agrees, within one standard deviation, with the SFR determined by \citet{robitaille10}, $0.68-1.45~M_{\sun}~\rm{yr}^{-1}$. These authors determine the star formation rate by counting intermediate to high-mass (3$M_{\sun}$ $\lesssim$ M $\lesssim$ 25$M_{\sun}$) pre-main sequence stars, in contrast to our use of a tracer of $\gtrsim$ 40$M_{\sun}$ stars, which are responsible for the bulk of the ionizing flux produced by young star clusters.

We note that SFCs inside the solar circle account for $80\%$ of the total galactic free-free luminosity, i.e., they are responsible for roughly $80\%$ of the star formation activity in the Milky Way galaxy.

\subsection{The relationship between free-free and $8\mu$m emission}
\label{ssec:ff8mrel}
\begin{figure}
\plotone{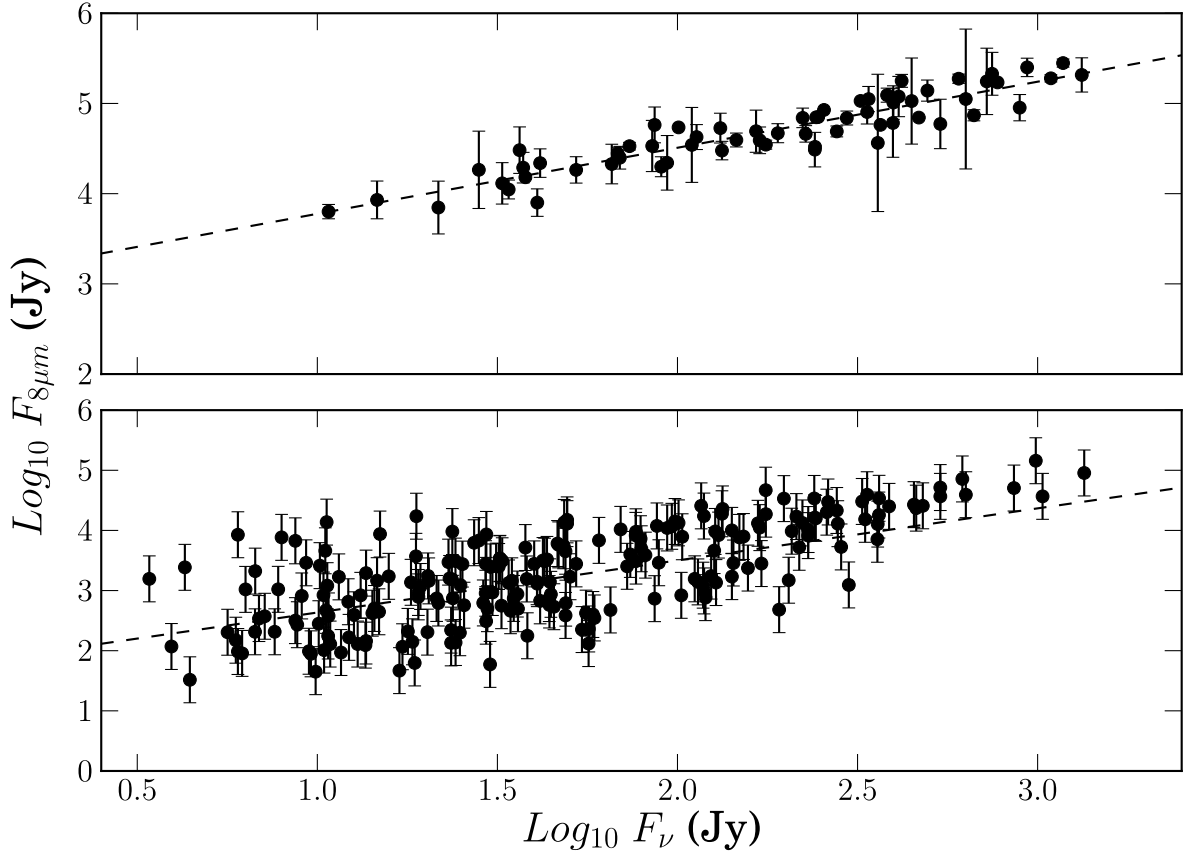}
\caption{\label{fig:flscl}Top: Log-log plot of 8$\mu$m vs. free-free flux of the WMAP sources with GLIMPSE or MSX coverage. The free-free flux of the WMAP sources is defined as the maximum between the isophotal and aperture-integrated 94 GHz free-free emission. The slope is found to be 0.73 $\pm$ 0.04 with a correlation coefficient 0.93 and reduced $\chi^{2}=4.87$ which shows that the two emissions are strongly correlated by approximately a linear relation. The error in the $F_{8 \mu m}$ is defined as the difference between the flux within SExtractor-determined aperture and the flux within the resized and reoriented aperture. Bottom: Log-log plot of 8$\mu$m vs. free-free flux of the SFCs within the WMAP sources with GLIMPSE or MSX coverage. The slope is found to be 0.87 $\pm$ 0.07 with a correlation coefficient 0.71 and reduced $\chi^{2}=3.44$ which shows a relatively weak correlation between the two emissions. The error is determined by taking the standard deviation of the total flux within apertures ranging from 0.8 to 2 times the original SFC radius; the mean error among the SFCs, $88\%$, is used.}
\end{figure}

With a more complete sample, we re-examine the relationship between WMAP free-free flux and 8$\mu$m from RM10 as shown in the top panel of Figure \ref{fig:flscl}:
\begin{equation}
\label{eq:ffscl}
F_{8 \mu m}~\propto~F^{0.73 \pm 0.04}_{\nu},
\end{equation}
where $F_{8 \mu m}$ is the aperture-integrated 8$\mu$m flux and $F_{\nu}$ is the aperture-integrated 94 GHz free-free emission from the WMAP free-free foreground emission map. The MSX 8$\mu$m flux is calibrated to GLIMPSE by an empirically determined relationship:
\begin{equation}
\label{eq:calib}
F_{GL} (Jy) = 5.19436~\pm~0.09625~F_{MSX} (Jy) - 5921.22~\pm~2587.89.
\end{equation}
The calibration is performed by first retrieving the total flux of overlapping regions between GLIMPSE and MSX images then fitting a linear line between the two fluxes. The correlation coefficient of the fit is 1.0.

\citet{calzetti} find the power law in Equation \ref{eq:ffscl} to be $0.94~\pm~0.02$ in the studies of extragalactic star formation, with the caveat that the analysis should be limited to the regions of star formation with a narrow spread in metallicity around the solar value. 

As~\citet{calzetti} note, 8$\mu$m emission traces the stellar light absorbed and re-emitted mostly by polycyclic aromatic hydrocarbon (PAH). Since both free-free and 8$\mu$m emission are tracers of star formation, a linear relationship is expected\footnote{Due to a coding error, RM10 reported that the $8\mu$m flux scaled as the square of the free-free flux.}.

The flux relationship for SFCs follows a similar power law, as shown in the bottom panel of Figure \ref{fig:flscl}:
\begin{equation}
\label{eq:ffscl_sfc}
F_{8 \mu m}~\propto~F^{0.87 \pm 0.07}_{\nu}.
\end{equation}

We have also found a tight correlation between the $8\mu$m emission and the radio recombination line flux in our SFCs: the correlation coefficient is $0.7$.

Finally, we note that the SFC 8$\mu$m flux exhibits a strong correlation with the projected size of the SFCs. 
Nearby sources of a fixed luminosity and physical size will appear both larger and have larger fluxes than equivalent sources at larger distances. If this were responsible for the correlation, one would expect that the flux would scale with the area on the sky (the surface brightness would be constant), whereas we see that the flux scales with the square root of the area.

A possible physical reason for such a correlation is the destruction of PAH molecules near the ionizing source~\citep{watson08}, reducing the 8$\mu$m emission in compact SFCs. MR10 also finds that the 8$\mu$m flux is dominated by emission at large projected radii rather than by the small, bright areas toward individual \ion{H}{2} regions. 

The sum of the 8$\mu$m emission inside SFCs, i.e., inside the bubble walls, only accounts for 33$\%$ of the $8\mu$m flux within the (more extended) WMAP sources. This suggests that the far UV (FUV) photons that excite PAH emission travel well past the bubble walls defining the SFCs. However, there may be sources other than central clusters that are responsible for the PAH emission in the diffuse region outside of the SFCs. 
%
One indication of this is that we find larger scatter in the 8$\mu$m flux when we sum over a  larger aperture in the GLIMPSE images. It is possible that nearby old B stars (with no association to SFCs) are emitting FUV photons. 

\subsection{SFC Expansion and Molecular Gas Turbulence}
\label{ssec:kinlum}
RM10 suggest that the primary driver of turbulence in Milky Way molecular clouds is the expanding bubbles seen inside SFCs. We test this idea using the more complete sample presented here.


Equation 3 of RM10 is used to determine the mechanical luminosity in the expansion of each SFC:
\begin{equation}
\label{eq:rm10_lmech}
L_{mech} = \frac{\pi}{2}\Sigma_{0}\Delta v_{c}^{3} r,
\end{equation}
where $\Sigma_{0} = 100 \pm 88 M_{\sun}~\rm{pc}^{-2}$ \citep{heyer09} is the surface density of a GMC (while their result shows an LTE-derived median surface density as $42~\pm~37~M_{\sun}~\rm{pc}^{-2}$, they consider this a lower limit and that the real value can change by a factor of 2-3 due to their neglect of abundance variations in the outer envelope of the clouds), $\Delta v_{c}$ corrected half-spread velocity, and r the geometric mean radius. For the SFCs without a measurement of the corrected expansion velocity, we use a mean expansion velocity of 11.7 $\pm$ 5.9 $\kms$. 

Limiting the analysis to SFCs inside the solar circle and correcting for the SFCs that do not have a KDA resolution as in $\S$~\ref{sssec:bias}, the total mechanical luminosity being injected into the interstellar medium (ISM) inside the solar circle is
\begin{equation}
\label{eq:lmech_sfc}
L_{mech} = 1.0 \pm 0.2~\rm{x}~10^{39}~\left(\frac{\Sigma_{0}}{100 M_{\sun} pc^{-2}}\right)~erg~s^{-1}
\end{equation}
where the error estimate results from propagation of the errors in the geometric radius, the expansion velocity, and the surface density. The contributions are 5$\%$ (radius), 17$\%$ (expansion velocity), 12$\%$ (surface density), and 1$\%$ from the MC simulation in $\S$ \ref{sssec:bias}. Once again, the most significant contributor is the error in the expansion velocity. 

This bubble kinetic luminosity should be compared to that required to maintain the velocity dispersion in the molecular gas within the solar circle:
\begin{equation}
\label{eq:lturb_basic}
L_{turb} = \frac{1}{2}\frac{Mv^{2}}{t_{0}},
\end{equation}
where M is the total molecular gas mass inside the solar circle, v molecular gas velocity, and $t_{0}$ the turbulence dissipation time scale defined by \citet{maclow99}:
\begin{equation}
\label{eq:maclow99_eq20}
t_{0} \simeq 1.0 \frac{\cal {L}}{v_{rms}}
\end{equation}
where $\cal {L}$ is the driving wavelength. We take the molecular gas velocity as the rms velocity and choose 2h, twice the scale height of the molecular disk, as the turbulence driving scale since it is the largest mode wavelength of a marginally gravitationally stable differentially rotating disk~\citep{goldreich_lynden_p2}. 

Using this, the turbulent luminosity is
\begin{equation}
\label{eq:lturb}
L_{turb} = \frac{1}{2}\frac{Mv^{3}}{2h}.
\end{equation}
We employ $h = 60 \pc$ \citep{malhotra}, $M$ = 1.0 x $10^{9}$ $M_{\sun}$, and $v = \sqrt{2\ln{2}}\sigma_{mol}$ with $\sigma_{mol}$ = 7 $\pm$ 1 $\kms$ \citep{malhotra}. The estimated solar circle molecular gas turbulent luminosity is then
\begin{equation}
\label{eq:lmech_gal}
L_{turb} = 1.5~\pm~0.6~\rm{x}~10^{39}~\left(\frac{\sigma}{7\kms}\right)^{3}~\left(\frac{60pc}{h}\right)~erg~s^{-1}
\end{equation}

The kinetic luminosity due to the expansion of the bubbles inside the SFCs can account for 65\% of the solar circle molecular gas turbulent luminosity. This suggests that the expansion of the bubbles inside SFCs is a major driver of turbulent motions of molecular gas. 

\section{DISCUSSION}
\label{sec:discuss}
We note that the bubble expansion velocity is larger than the Galaxy-wide average turbulent velocity, while the average bubble radius is smaller than the scale height of the molecular disk. It follows that the bubble time scale is significantly shorter than the turbulent decay time. Thus the total energy contained in the turbulent motions is significantly larger than that contained in the bubbles. It is also clear that the mass swept up in the bubble walls is significantly smaller than the total molecular gas mass inside the solar circle.

Estimates of the virial parameters of the most massive GMCs suggests that they are gravitationally bound objects \citep{heyer09}, yet the turbulence that supports them decays on a crossing time. Since the clouds are known not to collapse (such a collapse would lead to very high star formation rates) then either some way to sustain the turbulence is needed, or the GMCs must be disrupted. The fact that we find expanding bubbles in essentially all the rapidly star forming GMCs in the Milky Way strongly suggests that the clouds are disrupted rather than maintained in a quasi-steady state by injection of turbulent energy.

The bubble expansion can be driven by a number of possible mechanisms, including radiation pressure from star clusters in the bubble, warm ionized gas pressure, and hot (x-ray emitting) gas pressure~\citep{harper-clark, murray_rad10, lopez11, pellegrini}. It is possible that some of the low-luminosity SFCs may have low ionizing luminosities because they are $\gtrsim 4\Myr$ old, and hence may contain supernova remnants. In fact, a few SFCs are observed to contain known remnants and show bright emission in the WMAP synchrotron map. A uick visual inspection of WMAP W band synchrotron emission map shows that only about 23 SFCs ($8\%$) overlap bright synchrotron emission map. This suggests that supernovae do not disrupt GMCs. It also suggests, less directly, that supernovae may not be the dominant non-gravitational source of stirring in the ISM of the Milky Way. A more rigorous search for synchrotron emission in the faint, possibly older SFCs and WMAP sources would be very instructive.

When we say that bubble expansion is a major driver of turbulent motions, we have in mind the following picture. Imagine starting with a turbulently supported gas disk with scale height $h$, turbulent velocity $\sigma$, and a gas surface density $\Sigma$. Imagine further that the Toomre $Q=v_c\sigma/(\pi G\Sigma r)$ parameter is near unity, so that the disk is marginally unstable due to self-gravity ($v_c$ is the circular velocity of the galaxy, and $r$ is the galactocentric radius). Over time the turbulence decays, so that $\sigma$ decreases, as does $Q$. This triggers large scale instabilities, which drive turbulence on the scale $h$. However, some fraction of the gas collapses to form gravitationally bound objects, i.e., giant molecular clouds. These clouds contain subregions which are also gravitationally bound (often called clumps in the literature). These clumps are where the massive star clusters we study are formed. The star clusters emit radiation and stellar winds, which both act to inflate the bubbles we see. The bubbles expand rapidly, eventually disrupting the host GMC, and lifting gas above and below the galactic disk. Finally, the material falls back towards the disk plane, once again providing a source of large scale turbulence via gravitational instability.

\section{Conclusion}
\label{sec:concl}
We have analyzed 72 of the 88 Galactic WMAP sources and found 280 star forming complexes. Using 8$\mu$m fluxes, we assign free-free fluxes to all 280 SFCs. Of these 280 sources, 168 have a unique kinematic distance measurement. The 112 SFCs which have free-free flux estimates but which lack a unique kinematic distance measurement are accounted for by assuming a model of Galactic spiral arm, molecular hydrogen distribution and a detection sensitivity as a function of galactocentric radius. These 280 SFCs account for $74\%$ of the free-free flux of the Galaxy. The remaining $26\%$ of the free-free flux is shared between the flux associated with the sixteen WMAP sources lacking complete $8\mu$m coverage or bubbles, and the flux from outside the WMAP sources.

While our sample is biased against low luminosity sources at large heliocentric distances, we find that low luminosity sources have only a minor contribution to the total Galactic free-free luminosity. In fact we show the most luminous $9\%$ of all SFCs account for half the total Galactic free-free luminosity. The SFC luminosity function is described by two power-laws, with the slope in the high-luminosity tail consistent with values found in the literature on extragalactic \ion{H}{2} luminosity functions. 

With the SFC sample in hand, we extend the results of MR10 and RM10. The total galactic free-free luminosity is measured to be $1.6\pm0.3 \times 10^{27}{\rm erg\,}\s^{-1}\rm{\,Hz}^{-1}$, and the corresponding ionizing luminosity is $2.9\pm0.5\times10^{53}~{\rm photons\,}\s^{-1}$. We find a Galactic star formation rate of $1.2 \pm0.2M_{\sun}\yr^{-1}$, in agreement with MR10 and \citet{robitaille10}. We find a nearly linear relationship between 8$\mu$m and free-free flux of the WMAP sources and the SFCs, consistent with extragalactic studies. Solar circle SFCs are found to be responsible for 80\% of the total galactic free-free luminosity, and their kinetic energy injected per unit time into the ISM via expansion of the bubbles is equal to 65\% of the luminosity required to maintain the velocity dispersion seen in the solar circle molecular gas. This result supports the suggestion of RM10 that the expansion of the bubbles due to central massive clusters inside the SFCs is a major driver of the turbulent motion of the molecular gas in the inner Milky Way. 

\section{Acknowledgement}
\label{sec:ack}
We thank the anonymous referee, M.V.Kerkwijk, C.D.Matzner, and A.Shannon for valuable inputs and criticisms. This work is based (in part) on observations made with the Spizter Space Telescope, which is operated by the Jet Propulsion Laboratory, California Insitutte of Technology under a contract with NASA. This research has made use of the SIMBAD database, operated at CDS, Strasbourg, France. This research made use of data products from the Midcourse Space Experiment. Processing of the data was funded by the Ballistic Missile Defense Organization with additional support from the NASA Office of Space Science. This research made use of Montage, funded by the National Aeronautics and Space Administration's Earth Science Technology Office, Computation Technologies Project, under Cooperative Agreement Number NCC5-626 Between NASA and the California Institute of Technology. Montage is maintained by the NASA/IPAC Infrared Science Archive.

\begin{deluxetable}{ccccccccccccccccc}
\tablewidth{0pt} \rotate \tabletypesize{\scriptsize}
\tablecaption{Star Forming Complex Parameters \label{sfclist}}
\tablehead{ \colhead{No.} & \colhead{WMAP} & \colhead{$l$} &
\colhead{$b$} & \colhead{smaj} & \colhead{smin} &
\colhead{PA} & \colhead{$v_{LSR}$} & 
\colhead{$\Delta v_{m}$} & \colhead{$\Delta v_{c}$} & 
\colhead{$D$} & \colhead{$\sigma_{D}$} & \colhead{$<$R$>$} & \colhead{$f_{\nu}$} & 
\colhead{$t_{dyn}$} & \colhead{KDA} & \colhead{Ref.} \\
\colhead{} & \colhead{NAME} & \colhead{(deg)} &
\colhead{(deg)} & \colhead{(arcmin)} & \colhead{(arcmin)} &
\colhead{(deg)} & \colhead{(km s$^{-1}$)} & 
\colhead{(km s$^{-1}$)} & \colhead{(km s$^{-1}$)} & \colhead{(kpc)} & 
\colhead{(kpc)} & \colhead{(pc)} & \colhead{(Jy)} & 
\colhead{(Myr)} & \colhead{N/F} & \colhead{}}

\startdata
  0 & G359 &   0.129 &   0.044 &   4.6 &   3.4 & -30 &   -21.7\tablenotemark{b} & \nodata &    25.5\tablenotemark{c} &        8.0 &        2.4 &        9.2 &  127 &        0.4 & \nodata & \nodata \\
  1 & G359 &   0.165 &  -0.060 &   2.3 &   1.7 &  15 &    34.0\tablenotemark{d} & \nodata &    48.7\tablenotemark{d} &        8.0 &        2.4 &        4.6 &   34 &        0.1 & \nodata & \nodata \\
  2 & G359 &   0.285 &  -0.501 &  20.5 &  11.8 &  15 &    20.0 &     3.2 & \nodata &   9.0, 8.0 &   1.5, 1.5 &  40.6, 36. &  466 &   3.5, 3.1 & \nodata & \nodata \\
  3 & G359 &   0.502 &  -0.056 &   2.5 &   1.6 &  35 &    47.1 & \nodata & \nodata &   8.9, 8.1 &   1.5, 1.5 &   5.2, 4.8 &   31 &    \nodata & \nodata & \nodata \\
  4 & G359 &   0.510 &   0.183 &   1.4 &   1.0 &   0 & \nodata & \nodata & \nodata &    \nodata &    \nodata &    \nodata &    4 &    \nodata & \nodata & \nodata \\
  5 & G359 &   0.665 &   0.644 &   5.2 &   3.9 &  60 &     3.7 & \nodata & \nodata &  11.6, 5.4 &   7.5, 7.5 &  15.2, 7.1 &   25 &    \nodata & \nodata & \nodata \\
  6 & G359 &   0.829 &   0.207 &   2.4 &   1.3 & -65 &     9.7 & \nodata & \nodata &  10.5, 6.5 &   2.6, 2.6 &   5.4, 3.4 &    6 &    \nodata & \nodata & \nodata \\
  7 & G359 &   1.137 &  -0.087 &   3.1 &   2.1 &  15 &   -21.7 & \nodata & \nodata &    \nodata &    \nodata &    \nodata &   21 &    \nodata & \nodata & \nodata \\
  8 & G359 &   1.341 &   0.118 &   2.7 &   2.4 &  50 &   -12.0 & \nodata & \nodata &       12.0 &        3.6 &        8.9 &   13 &    \nodata & \nodata & \nodata \\
  9 &   G3 &   2.268 &   0.234 &   3.6 &   2.4 & -10 &     4.9 & \nodata & \nodata &  14.0, 4.0 &   1.2, 1.2 &  12.0, 3.4 &    6 &    \nodata & \nodata & \nodata \\
 10 &   G3 &   2.515 &  -0.032 &   1.7 &   1.3 &  40 &     8.3 & \nodata & \nodata &  13.0, 4.0 &   3.7, 3.7 &   5.5, 1.7 &    1 &    \nodata & \nodata & \nodata \\
 11 &   G3 &   2.602 &   0.162 &   2.4 &   1.6 & -70 &   102.4 & \nodata & \nodata &   9.2, 7.8 &   0.1, 0.1 &   5.3, 4.5 &    3 &    \nodata & \nodata & \nodata \\
 12 &   G3 &   2.866 &  -0.032 &   5.1 &   3.2 & -40 &    -1.8 & \nodata & \nodata &       12.0 &        3.6 &       14.1 &   14 &    \nodata & \nodata & \nodata \\
 13 &   G3 &   3.297 &  -0.040 &   7.2 &   4.8 & -40 &     6.0 &     4.0 & \nodata &  12.0, 2.6 &   3.6, 3.6 &  20.6, 4.5 &   32 &   1.8, 0.4 & \nodata & \nodata \\
 14 &   G3 &   3.656 &  -0.113 &   1.5 &   1.0 & -40 &     4.6 & \nodata & \nodata &  12.0, 1.9 &   3.6, 3.6 &   4.4, 0.7 &    1 &    \nodata & \nodata & \nodata \\
\enddata
\tablenotetext{b}{This SFC selection is motivated by \citet{cotera05} and \citet{lang01}. $v_{LSR}$ is determined by taking the average between filaments E1, E2 and the Arches cluster velocity~\citep{cotera05}.}
\tablenotetext{c}{The expansion velocity is determined by taking the mean FWHM velocity among E2, W1, W2 filaments~\citep{cotera05, lang01}.}
\tablenotetext{d}{Velocities are taken from~\citet{lang97}.}
\tablerefs{(1) \citet{kda_BUcat}; (2) \citet{kda_busfield}; (3) \citet{kda_sewilo}; (4) \citet{kda_fish}; (5) \citet{kda_kolpak}; (6) \citet{kda_palagi}; (7) \citet{vrad_caswell_haynes}; (8) \citet{vrad_gill}}
\tablecomments{The KDA Ref. indicates the literatures used to resolve kinematic distance ambiguity. In cases where no reference is given, the kinematic distance is either unique, undetermined, or we present both the near and far distances.}
\tablecomments{(This table is available in its entirety in a machine-readble form in the online journal. A portion is shown here for guidance regarding its form and content.)}
\end{deluxetable}

\begin{deluxetable}{cccccc}
\tablewidth{0pt} \tabletypesize{\footnotesize}
\tablecaption{HII Region RRL Velocities \label{hiilist}}
\tablehead{ \colhead{SFC} & \colhead{Name} &
\colhead{$l$} & \colhead{$b$} &
\colhead{$v_{LSR}$} & 
\colhead{Ref.}}

\startdata
  2 &           [KC97c] G000.3-00.5  &   0.2840 &  -0.4781 &   20.0 &  5 \\
  2 &           [KC97c] G000.4-00.5  &   0.3940 &  -0.5400 &   24.0 &  5 \\
  2 &           [KC97c] G000.5-00.7  &   0.4890 &  -0.6681 &   17.5 &  5 \\
  2 &        MSX6C G000.5799-00.6302 &   0.5718 &  -0.6279 &   20.0 &  5 \\
  3 &              GAL 000.49-00.05  &   0.4905 &  -0.0600 &   45.8 & 13 \\
  3 &          [WWB83] G000.51-0.04  &   0.5137 &  -0.0394 &   47.1 & 13 \\
  3 &           [KC97c] G000.5-00.1  &   0.5180 &  -0.0647 &   47.1 &  5 \\
  5 &         [LPH96] 000.640+0.623  &   0.6403 &   0.6239 &    3.7 &  7 \\
  6 &           [KC97c] G000.8+00.2  &   0.8290 &   0.1934 &    9.7 &  5 \\
  7 &           [KC97c] G001.1-00.1  &   1.1488 &  -0.0618 &  -21.7 &  5 \\
  8 &              GAL 001.32+00.09  &   1.3228 &   0.0860 &  -12.0 &  1 \\
  9 &           [KC97c] G002.3+00.2  &   2.3031 &   0.2430 &    4.9 &  6 \\
 10 &         [LPH96] 002.510-0.028  &   2.5099 &  -0.0281 &    8.3 &  7 \\
 11 &           [KC97c] G002.6+00.1  &   2.6110 &   0.1351 &  102.4 &  6 \\
 11 &               IRAS 17480-2636  &   2.6130 &   0.1340 &  102.4 &  6 \\
\enddata
\tablecomments{References: (1)\citet{vrad_caswell_haynes}; (2)\citet{vrad_heiles}; (3)\citet{vrad_kim}; (4)\citet{vrad_kuchar} 87GB; (5)\citet{vrad_kuchar} PMN; (6)\citet{vrad_lockman}; (7)\citet{vrad_lockman96}; (8)\citet{vrad_paladini}; (9)\citet{vrad_quireza}; (10)\citet{vrad_wilson}; (11)\citet{vrad_caswell72}; (12)\citet{vrad_dieter}; (13)\citet{wink83}; (14)\citet{peeters02}; (15)\citet{araya07}; (16)\citet{russeil05} \\ \\ (This table is available in its entirety in a machine-readble form in the online journal. A portion is shown here for guidance regarding its form and content.)}
\end{deluxetable}

\begin{deluxetable}{ccccccccc}
\tablewidth{0pt} \rotate \tabletypesize{\footnotesize}
\tablecaption{Star Forming Complex Comparison \label{sfccomp}}
\tablehead{ \colhead{} & \colhead{$l$} & \colhead{$b$} &
\colhead{dsmaj} & \colhead{dsmin} &
\colhead{$dv_{LSR}$} & \colhead{$d\Delta v_{m}$} &
\colhead{$d\Delta v_{c}$} & \colhead{$dD_{kin}$} \\
\colhead{} & \colhead{(deg)} & \colhead{(deg)} & 
\colhead{(\%)} & \colhead{(\%)} & \colhead{(\%)} &
\colhead{(\%)} & \colhead{(\%)} & \colhead{(\%)}}

\startdata
Type 1 &  -0.04 $\pm$   0.14 &   0.10 $\pm$   0.16 &  19.77 $\pm$  23.11 &  17.51 $\pm$  23.25 &  14.39 $\pm$  30.35 &  18.96 $\pm$  20.29 &  22.50 $\pm$  13.93 &  16.09 $\pm$  37.76 \\
Type 2 &  -0.19 $\pm$   0.30 &  -0.13 $\pm$   0.21 & 327.33 $\pm$ 338.95 & 328.95 $\pm$ 327.11 &   6.58 $\pm$   7.26 &  75.69 $\pm$  27.78 &  61.01 $\pm$  33.67 &  41.77 $\pm$  34.59 \\
   All &  -0.07 $\pm$   0.17 &   0.04 $\pm$   0.14 &  67.79 $\pm$ 129.19 &  70.89 $\pm$ 156.38 &  29.14 $\pm$  57.76 &  28.32 $\pm$  21.19 &  27.33 $\pm$  19.43 &  22.13 $\pm$  32.43 \\
\enddata
\end{deluxetable}

\begin{deluxetable}{cccc}
\tablewidth{0pt} \rotate \tabletypesize{\footnotesize}
\tablecaption{Monte Carlo Simulation Result \label{sfcsimres}}
\tablehead{ \colhead{} & \colhead{Data} & \colhead{Blind-MC} & \colhead{Data-MC}}

\startdata
   $L_{ALL,27}$ & 1.15 $\pm$ 0.15 & 1.13 $\pm$ 0.28 & 1.27 $\pm$ 0.02\\
    $L_{SC,27}$ & 0.96 $\pm$ 0.12 & 0.96 $\pm$ 0.26 & 1.16 $\pm$ 0.02\\
  $L_{NEAR,27}$ & 0.41 $\pm$ 0.09 & 0.40 $\pm$ 0.12 & 0.55 $\pm$ 0.02\\
   $L_{FAR,27}$ & 0.55 $\pm$ 0.08 & 0.56 $\pm$ 0.24 & 0.62 $\pm$ 0.03\\
\enddata
\tablecomments{$L_{ALL}$ of Data-MC calculation is performed with correcting the OSC luminosity by the overestimate of kinematic distance as mentioned in Section \ref{sssec:vel_dist}. This correction decreases the luminosity by a $\sim$ 12$\%$.}
\end{deluxetable}

\bibliography{sfckin}

\begin{thebibliography}{72}
\expandafter\ifx\csname natexlab\endcsname\relax\def\natexlab#1{#1}\fi

\bibitem[{{Anderson} \& {Bania}(2009)}]{kda_BUcat}
{Anderson}, L.~D., \& {Bania}, T.~M. 2009, \apj, 690, 706

\bibitem[{{Araya} {et~al.}(2007){Araya}, {Hofner}, {Goss}, {et~al.}}]{araya07}
{Araya}, E., {Hofner}, P., {Goss}, W.~M., {et~al.} 2007, \apjs, 170, 152

\bibitem[{{Bennett} {et~al.}(2003){Bennett}, {Hill}, {Hinshaw},
  {et~al.}}]{wmap_fg}
{Bennett}, C.~L., {Hill}, R.~S., {Hinshaw}, G., {et~al.} 2003, \apjs, 148, 97

\bibitem[{{Binney} \& {Merrifield}(1998)}]{binney98}
{Binney}, J., \& {Merrifield}, M. 1998, {Galactic Astronomy}, ed. {Binney,
  J.~\& Merrifield, M.}

\bibitem[{{Busfield} {et~al.}(2006){Busfield}, {Purcell}, {Hoare},
  {et~al.}}]{kda_busfield}
{Busfield}, A.~L., {Purcell}, C.~R., {Hoare}, M.~G., {et~al.} 2006, \mnras,
  366, 1096

\bibitem[{{Calzetti} {et~al.}(2007){Calzetti}, {Kennicutt}, {Engelbracht},
  {et~al.}}]{calzetti}
{Calzetti}, D., {Kennicutt}, R.~C., {Engelbracht}, C.~W., {et~al.} 2007, \apj,
  666, 870

\bibitem[{{Cappa} {et~al.}(2011){Cappa}, {Barb{\'a}}, {Duronea},
  {et~al.}}]{cappa11}
{Cappa}, C.~E., {Barb{\'a}}, R., {Duronea}, N.~U., {et~al.} 2011, \mnras, 415,
  2844

\bibitem[{{Caswell}(1972)}]{vrad_caswell72}
{Caswell}, J.~L. 1972, Australian Journal of Physics, 25, 443

\bibitem[{{Caswell} \& {Haynes}(1987)}]{vrad_caswell_haynes}
{Caswell}, J.~L., \& {Haynes}, R.~F. 1987, \aap, 171, 261

\bibitem[{{Clemens}(1985)}]{clemens85}
{Clemens}, D.~P. 1985, \apj, 295, 422

\bibitem[{{Cotera} {et~al.}(2005){Cotera}, {Colgan}, {Simpson}, \&
  {Rubin}}]{cotera05}
{Cotera}, A.~S., {Colgan}, S.~W.~J., {Simpson}, J.~P., \& {Rubin}, R.~H. 2005,
  \apj, 622, 333

\bibitem[{{Dame} {et~al.}(2001){Dame}, {Hartmann}, \& {Thaddeus}}]{dame01}
{Dame}, T.~M., {Hartmann}, D., \& {Thaddeus}, P. 2001, \apj, 547, 792

\bibitem[{{Dame} {et~al.}(1987){Dame}, {Ungerechts}, {Cohen},
  {et~al.}}]{dame87}
{Dame}, T.~M., {Ungerechts}, H., {Cohen}, R.~S., {et~al.} 1987, \apj, 322, 706

\bibitem[{{Dieter}(1967)}]{vrad_dieter}
{Dieter}, N.~H. 1967, \apj, 150, 435

\bibitem[{{Downes} {et~al.}(1980){Downes}, {Wilson}, {Bieging}, \&
  {Wink}}]{kda_downes}
{Downes}, D., {Wilson}, T.~L., {Bieging}, J., \& {Wink}, J. 1980, \aaps, 40,
  379

\bibitem[{{Fish} {et~al.}(2003){Fish}, {Reid}, {Wilner}, \&
  {Churchwell}}]{kda_fish}
{Fish}, V.~L., {Reid}, M.~J., {Wilner}, D.~J., \& {Churchwell}, E. 2003, \apj,
  587, 701

\bibitem[{{Gillespie} {et~al.}(1977){Gillespie}, {Huggins}, {Sollner},
  {et~al.}}]{vrad_gill}
{Gillespie}, A.~R., {Huggins}, P.~J., {Sollner}, T.~C.~L.~G., {et~al.} 1977,
  \aap, 60, 221

\bibitem[{{Goldreich} \& {Kwan}(1974)}]{1974ApJ...189..441G}
{Goldreich}, P., \& {Kwan}, J. 1974, \apj, 189, 441

\bibitem[{{Goldreich} \& {Lynden-Bell}(1965)}]{goldreich_lynden_p2}
{Goldreich}, P., \& {Lynden-Bell}, D. 1965, \mnras, 130, 125

\bibitem[{{G{\'o}mez}(2006)}]{gomez06}
{G{\'o}mez}, G.~C. 2006, \aj, 132, 2376

\bibitem[{{Gonzalez Delgado} \& {Perez}(1997)}]{delgado97}
{Gonzalez Delgado}, R.~M., \& {Perez}, E. 1997, \apjs, 108, 199

\bibitem[{{Harper-Clark} \& {Murray}(2009)}]{harper-clark}
{Harper-Clark}, E., \& {Murray}, N. 2009, \apj, 700, 137

\bibitem[{{Heiles} {et~al.}(1996){Heiles}, {Reach}, \& {Koo}}]{vrad_heiles}
{Heiles}, C., {Reach}, W.~T., \& {Koo}, B. 1996, \apj, 466, 191

\bibitem[{{Heyer} {et~al.}(2009){Heyer}, {Krawczyk}, {Duval}, \&
  {Jackson}}]{heyer09}
{Heyer}, M., {Krawczyk}, C., {Duval}, J., \& {Jackson}, J.~M. 2009, \apj, 699,
  1092

\bibitem[{{Karzas} \& {Latter}(1961)}]{karzas61}
{Karzas}, W.~J., \& {Latter}, R. 1961, \apjs, 6, 167

\bibitem[{{Kennicutt} {et~al.}(1989){Kennicutt}, {Edgar}, \& {Hodge}}]{KEH}
{Kennicutt}, Jr., R.~C., {Edgar}, B.~K., \& {Hodge}, P.~W. 1989, \apj, 337, 761

\bibitem[{{Kim} \& {Koo}(2001)}]{vrad_kim}
{Kim}, K., \& {Koo}, B. 2001, \apj, 549, 979

\bibitem[{{Klessen} \& {Hennebelle}(2010)}]{klessen10}
{Klessen}, R.~S., \& {Hennebelle}, P. 2010, \aap, 520, A17

\bibitem[{{Kolpak} {et~al.}(2003){Kolpak}, {Jackson}, {Bania}, {Clemens}, \&
  {Dickey}}]{kda_kolpak}
{Kolpak}, M.~A., {Jackson}, J.~M., {Bania}, T.~M., {Clemens}, D.~P., \&
  {Dickey}, J.~M. 2003, \apj, 582, 756

\bibitem[{{Kuchar} \& {Clark}(1997)}]{vrad_kuchar}
{Kuchar}, T.~A., \& {Clark}, F.~O. 1997, \apj, 488, 224

\bibitem[{{Lang} {et~al.}(2001){Lang}, {Goss}, \& {Morris}}]{lang01}
{Lang}, C.~C., {Goss}, W.~M., \& {Morris}, M. 2001, \aj, 121, 2681

\bibitem[{{Lang} {et~al.}(1997){Lang}, {Goss}, \& {Wood}}]{lang97}
{Lang}, C.~C., {Goss}, W.~M., \& {Wood}, D.~O.~S. 1997, \apj, 474, 275

\bibitem[{{Li} {et~al.}(2005){Li}, {Mac Low}, \& {Klessen}}]{li05}
{Li}, Y., {Mac Low}, M.-M., \& {Klessen}, R.~S. 2005, \apjl, 620, L19

\bibitem[{{Li} \& {Nakamura}(2006)}]{li06}
{Li}, Z.-Y., \& {Nakamura}, F. 2006, \apjl, 640, L187

\bibitem[{{Lockman}(1989)}]{vrad_lockman}
{Lockman}, F.~J. 1989, \apjs, 71, 469

\bibitem[{{Lockman} {et~al.}(1996){Lockman}, {Pisano}, \&
  {Howard}}]{vrad_lockman96}
{Lockman}, F.~J., {Pisano}, D.~J., \& {Howard}, G.~J. 1996, \apj, 472, 173

\bibitem[{{Lopez} {et~al.}(2011){Lopez}, {Krumholz}, {Bolatto}, {Prochaska}, \&
  {Ramirez-Ruiz}}]{lopez11}
{Lopez}, L.~A., {Krumholz}, M.~R., {Bolatto}, A.~D., {Prochaska}, J.~X., \&
  {Ramirez-Ruiz}, E. 2011, \apj, 731, 91

\bibitem[{{Mac Low} \& {Klessen}(2004)}]{maclow04}
{Mac Low}, M., \& {Klessen}, R.~S. 2004, Reviews of Modern Physics, 76, 125

\bibitem[{{Mac Low} {et~al.}(1998){Mac Low}, {Klessen}, {Burkert}, \&
  {Smith}}]{maclow}
{Mac Low}, M., {Klessen}, R.~S., {Burkert}, A., \& {Smith}, M.~D. 1998,
  Physical Review Letters, 80, 2754

\bibitem[{{Mac Low}(1999)}]{maclow99}
{Mac Low}, M.-M. 1999, \apj, 524, 169

\bibitem[{{Malhotra}(1994)}]{malhotra}
{Malhotra}, S. 1994, \apj, 433, 687

\bibitem[{{Matzner}(2002)}]{matzner02}
{Matzner}, C.~D. 2002, \apj, 566, 302

\bibitem[{{McKee} \& {Williams}(1997)}]{mckee}
{McKee}, C.~F., \& {Williams}, J.~P. 1997, \apj, 476, 144

\bibitem[{{McMillan} \& {Binney}(2010)}]{mcmillan10}
{McMillan}, P.~J., \& {Binney}, J.~J. 2010, \mnras, 402, 934

\bibitem[{{Mezger}(1978)}]{mezger}
{Mezger}, P.~O. 1978, \aap, 70, 565

\bibitem[{{Murray}(2011)}]{gmc}
{Murray}, N. 2011, \apj, 729, 133

\bibitem[{{Murray} {et~al.}(2005){Murray}, {Quataert}, \& {Thompson}}]{mqt05}
{Murray}, N., {Quataert}, E., \& {Thompson}, T.~A. 2005, \apj, 618, 569

\bibitem[{{Murray} {et~al.}(2010){Murray}, {Quataert}, \&
  {Thompson}}]{murray_rad10}
---. 2010, \apj, 709, 191

\bibitem[{{Murray} \& {Rahman}(2010)}]{paper1}
{Murray}, N., \& {Rahman}, M. 2010, \apj, 709, 424

\bibitem[{{Osterbrock}(1989)}]{osterbrock89}
{Osterbrock}, D.~E. 1989, {Astrophysics of gaseous nebulae and active galactic
  nuclei}, ed. {Osterbrock, D.~E.}

\bibitem[{{Paladini} {et~al.}(2003){Paladini}, {Burigana}, {Davies},
  {et~al.}}]{vrad_paladini}
{Paladini}, R., {Burigana}, C., {Davies}, R.~D., {et~al.} 2003, \aap, 397, 213

\bibitem[{{Palagi} {et~al.}(1993){Palagi}, {Cesaroni}, {Comoretto}, {Felli}, \&
  {Natale}}]{kda_palagi}
{Palagi}, F., {Cesaroni}, R., {Comoretto}, G., {Felli}, M., \& {Natale}, V.
  1993, \aaps, 101, 153

\bibitem[{{Peeters} {et~al.}(2002){Peeters}, {Mart{\'{\i}}n-Hern{\'a}ndez},
  {Damour}, {et~al.}}]{peeters02}
{Peeters}, E., {Mart{\'{\i}}n-Hern{\'a}ndez}, N.~L., {Damour}, F., {et~al.}
  2002, \aap, 381, 571

\bibitem[{{Pellegrini} {et~al.}(2011){Pellegrini}, {Baldwin}, \&
  {Ferland}}]{pellegrini}
{Pellegrini}, E.~W., {Baldwin}, J.~A., \& {Ferland}, G.~J. 2011, \apj, 738, 34

\bibitem[{{Quireza} {et~al.}(2006){Quireza}, {Rood}, {Balser}, \&
  {Bania}}]{vrad_quireza}
{Quireza}, C., {Rood}, R.~T., {Balser}, D.~S., \& {Bania}, T.~M. 2006, \apjs,
  165, 338

\bibitem[{{Rahman} {et~al.}(2011{\natexlab{a}}){Rahman}, {Matzner}, \&
  {Moon}}]{candidate}
{Rahman}, M., {Matzner}, C., \& {Moon}, D.-S. 2011{\natexlab{a}}, \apjl, 728,
  L37

\bibitem[{{Rahman} {et~al.}(2011{\natexlab{b}}){Rahman}, {Moon}, \&
  {Matzner}}]{rahman11}
{Rahman}, M., {Moon}, D.-S., \& {Matzner}, C.~D. 2011{\natexlab{b}}, \apjl,
  743, L28

\bibitem[{{Rahman} \& {Murray}(2010)}]{paper2}
{Rahman}, M., \& {Murray}, N. 2010, \apj, 719, 1104

\bibitem[{{Reid} {et~al.}(2009){Reid}, {Menten}, {Zheng}, {et~al.}}]{reid09}
{Reid}, M.~J., {Menten}, K.~M., {Zheng}, X.~W., {et~al.} 2009, \apj, 700, 137

\bibitem[{{Robitaille} \& {Whitney}(2010)}]{robitaille10}
{Robitaille}, T.~P., \& {Whitney}, B.~A. 2010, \apjl, 710, L11

\bibitem[{{Russeil} {et~al.}(2005){Russeil}, {Adami}, {Amram},
  {et~al.}}]{russeil05}
{Russeil}, D., {Adami}, C., {Amram}, P., {et~al.} 2005, \aap, 429, 497

\bibitem[{{Salpeter}(1955)}]{salpeter55}
{Salpeter}, E.~E. 1955, \apj, 121, 161

\bibitem[{{Sellwood} \& {Balbus}(1999)}]{sellwood99}
{Sellwood}, J.~A., \& {Balbus}, S.~A. 1999, \apj, 511, 660

\bibitem[{{Sewilo} {et~al.}(2004){Sewilo}, {Watson}, {Araya},
  {et~al.}}]{kda_sewilo}
{Sewilo}, M., {Watson}, C., {Araya}, E., {et~al.} 2004, \apjs, 154, 553

\bibitem[{{Shu} {et~al.}(1987){Shu}, {Adams}, \& {Lizano}}]{shu}
{Shu}, F.~H., {Adams}, F.~C., \& {Lizano}, S. 1987, \araa, 25, 23

\bibitem[{{Solomon} {et~al.}(1987){Solomon}, {Rivolo}, {Barrett}, \&
  {Yahil}}]{solomon87}
{Solomon}, P.~M., {Rivolo}, A.~R., {Barrett}, J., \& {Yahil}, A. 1987, \apj,
  319, 730

\bibitem[{{Sutherland}(1998)}]{sutherland}
{Sutherland}, R.~S. 1998, \mnras, 300, 321

\bibitem[{{Taylor} \& {Cordes}(1993)}]{taylor93}
{Taylor}, J.~H., \& {Cordes}, J.~M. 1993, \apj, 411, 674

\bibitem[{{Wang} {et~al.}(2010){Wang}, {Li}, {Abel}, \& {Nakamura}}]{wang10}
{Wang}, P., {Li}, Z.-Y., {Abel}, T., \& {Nakamura}, F. 2010, \apj, 709, 27

\bibitem[{{Watson} {et~al.}(2008){Watson}, {Povich}, {Churchwell},
  {et~al.}}]{watson08}
{Watson}, C., {Povich}, M.~S., {Churchwell}, E.~B., {et~al.} 2008, \apj, 681,
  1341

\bibitem[{{Wilson} {et~al.}(1970){Wilson}, {Mezger}, {Gardner}, \&
  {Milne}}]{vrad_wilson}
{Wilson}, T.~L., {Mezger}, P.~G., {Gardner}, F.~F., \& {Milne}, D.~K. 1970,
  \aap, 6, 364

\bibitem[{{Wink} {et~al.}(1983){Wink}, {Wilson}, \& {Bieging}}]{wink83}
{Wink}, J.~E., {Wilson}, T.~L., \& {Bieging}, J.~H. 1983, \aap, 127, 211

\end{thebibliography}

\end{document}